# Elementary conjugated fragments in acyclic polyenes with heteroatoms

Viktorija Gineityte

**Abstract**
The study is aimed at revealing the most important substructures (fragments) of polyenes with heteroatom(s) determining the alteration in the conjugation energy (CE) of the whole compound due to substitution and the relevant charge redistribution. The systems are modelled as sets of weakly-interacting formally-double bonds, where the formally-single bonds represent the interaction. Expressions for total energies and populations of basis orbitals are then derived in the form of power series with respect to two small parameters, viz. the averaged difference in Coulomb integrals of $2p_z$ AOs of heteroatoms from those of carbon atoms and the mean value of resonance parameters of formally-single bonds. Analysis of these series shows that conjugated substructures (I–L) consisting of two connected formally-double bonds (I and L) and containing at least s single heteroatom (X) play the role of fragments being sought. Nine potential fragments of this type are considered separately that differ one from another in the number of heteroatoms and/or their relative positions inside and are called elementary conjugated fragments. On the basis of expressions for contributions of the latter to characteristics of the whole compound, these fragments are subsequently classified into two groups, viz. (A) those contributing to growth of the overall CE (i.e. stabilization of the system) and participating in the charge redistribution among formally-double bonds (e.g. X=C–C=C, X=C–X=C, etc) and (B) destabilizing fragments characterized by zero increments to the same charge transfer (e.g. C=X–C=C, X=C–C=X, etc). This classification of elementary fragments is rationalized in terms of local (direct and indirect) interactions of orbitals of formally-double bonds.

## Introduction

Various experimental and computational methods usually yield characteristics of molecules in the form of bare numbers. The main task of any theory is then to relate these numbers to particular aspects of constitution of the compounds concerned. Obviously, rules embracing entire series or classes of similar compounds are highly desirable here. Moreover, additive forms of these structure/property relations are preferable over more sophisticated ones.

Simple models of structures often prove to be helpful in achieving the above-specified ends. Choice of a particular model, however, depends on systems under interest. For example, graphs representing the carbon backbones along with their adjacency matrices (AMs) [1-3] are most commonly used as models of conjugated hydrocarbons. Moreover, proportionality of these AMs to the relevant Hückel type Hamiltonian matrices [3] enables one to study the dependences of electronic structure characteristics of these hydrocarbons (especially of their total $\pi$–electron energies) upon parameters (descriptors) of respective graphs (see e.g. [4, 5]). An important point here also is that numerous options of potential descriptors still remain even after specifying the model. The reason possibly lies in the absence of a unique definition of the structure itself [6].

Two alternative perspectives on structure/property relations reveal themselves from analysis of the overall diversity of the relevant contributions including those devoted to conjugated hydrocarbons. The first one is based on employment of descriptors (parameters) of the whole graph and may be accordingly called the global perspective. Simple characteristics of graphs (e.g. numbers of vertices and edges [4], the mean length (of conjugation) [7, 8], etc) may be mentioned here as examples along with more sophisticated descriptors of global nature, such as the Z index [7-9]. Various intermediate criteria (i.e. those depending on the overall constitution of the graph concerned indirectly) also often are invoked, e.g. the number of Kekulé structures [4] along with that of the Dewar resonance structures [10, 11], as well as numbers of conjugated circuits of different size [12].

The opposite (i.e. the local) perspective consists in search for substructures (fragments) of the given system (and/or for subgraphs of the relevant graph) that determine the property under interest. Decomposition of total $\pi$-electron energies of (poly)cyclic conjugated hydrocarbons into increments of individual cycles and their pairs [13] may be referred to here. At the same time, analogous energies of acyclic polyenes were shown to take an approximately additive form with respect to contributions of individual branchings of the chain [14]. Again, select eigenvalues of AMs of some conjugated hydrocarbons have been traced back to subspectral fragments (subgraphs) embedded in the given graph [15-17].

Heteroconjugated molecules compose an even more extended and important class of chemical compounds. In respect of modelling, however, these systems offer us a more complicated object. In accordance with the above-discussed graph-theoretical tradition, these molecules are modelled by somewhat generalized graphs [18-21] containing self-loops (weighted vertices) at the site(s) of heteroatom(s). It is then no surprise that the overall technique becomes more involved and cumbersome, especially in the case of more than one heteroatom [21]. Nevertheless, relative stabilities of alternative derivatives proved to be determined by global descriptors of the standard graph of the respective parent hydrocarbons, viz. by the so-called topological index [18] and/or the Dewar index [19] at the site of substitution. Given that the weight of vertices associated with heteroatom(s) is assumed to be a small quantity, the approach under discussion resolves itself into the well-known classical perturbation theory of Coulson and Longuet-Higgins (abbreviated below as CLHPT) [22, 23], where heteroatom(s) are represented by small alterations in Coulomb integrals(s) of the standard Hückel model and self- and mutual polarizabilities of atoms of the parent hydrocarbon play the role of the principal quantities. Moreover, a fortunate combination of the CLHPT with a definite graph-theoretical technique resulted into rules representing the extinction rates of mutual polarizabilities ($\pi_{rs}$) with increasing distances between the *r*th and *s*th carbon atoms of an acyclic polyene [24]. It also deserves adding here that polarizabilities are expressible via coefficients of respective delocalized (canonical) molecular orbitals (MOs) [22] and also belong to global characteristics of the parent hydrocarbon. Meanwhile, the local perspective on heteroatom influence appearently has not yet been realized. Moreover, a question arises whether the molecular graph itself is the most appropriate model for achieving this end.

In this context, a clear distinction should be made between (poly)cyclic (aromatic) and acyclic hydrocarbons. Indeed, employment of the molecular graph as a model is based on an assumption that carbon-carbon bonds are uniform (or almost uniform) over the whole backbone of the conjugated hydrocarbon concerned. Such a requirement is more or less met for (poly)cyclic (aromatic) hydrocarbons, particularly for benzenoids (see e.g. [25]). Meanwhile, the analogous bond lengths are far from being uniform over chains of acyclic polyenes, especially of small- and medium-sized ones. The point is that formally-double (C=C) bonds still remain significantly shorter here as compared to formally-single (C–C) ones, although these are somewhat lengthened and shortened, respectively, vs. their standard values [10, 26-

28] (c.f. the so-called bond length alternation (BLA) [26, 28]). Consequently, the molecular graph and/or its AM is merely one of two possible approximate models of acyclic polyenes. Another alternative consists in consideration of an acyclic $\pi$-electron system as a set of weakly-interacting initially-double (C=C) bonds, where the single (C–C) bonds play the role of the interaction. Such a perturbational treatment of the carbon backbone itself enabled us previously [29, 30] to express the relevant total $\pi$-electron energy in the form of power series with respect to a single parameter ($\gamma$) coinciding with the ratio of the averaged resonance parameter of single bonds to that of double bonds. Moreover, separate members of these series proved to be representable via local (direct and indirect) interactions between orbitals of double (C=C) bonds. Finally, the decisive contribution to the overall conjugation energy (CE) of an acyclic polyene coincided with the sum of local increments of individual butadiene-like (C=C–C=C) fragments. On this basis, one can expect good prospects of this perturbative approach for realization of the local perspective on heteroatom influence.

In the present study, an attempt is undertaken to extend the above-described achievements to acyclic polyenes with any number of heteroatoms irrespective of their positions inside the parent hydrocarbon. In this connection, we are about to introduce an additional perturbational parameter ($\alpha$) into the former model of the parent polyene that coincides with the averaged alteration in the relevant Coulomb integral(s) of carbon atom(s) after replacing them by heteroatom(s). The main aim of the study consists in derivation of the relevant dual series and its analysis in order to reveal the most important fragments responsible for the alteration in the overall CE due to substitution and for the relevant charge redistribution. Evaluations of relative increments of different fragments (if any) to characteristics of the whole compound also is among our aims.

## Results and discussion

*General results for any polyene with heteroatom(s)*

Let us start with a certain parent polyne ($C_{2N}H_{2N+2}$) containing *N* formally-double (C=C) bonds. At the Hückel level [31, 32], the $\pi$-electron system of this molecule is representable by the 2*N*-dimensional basis of $2p_z$ atomic orbitals (AOs) of carbon atoms { $\chi$ }. Let these AOs be characterized by uniform Coulomb parameters coinciding with the energy reference point and consequently taking a zero value as usual. Resonance parameters between AOs inside formally-double bonds also are assumed to be uniform and to coincide with the (negative) energy unit in addition. Meanwhile, the averaged resonance parameter of formally-single (C–C) bonds will be denoted by $\gamma$ and supposed to be a small (first order) term vs. the above-specified energy unit (Note that $\gamma > 0$). Finally, resonance parameters for pairs of more remote AOs will be ignored [31, 32]. Since our polyene belongs to alternant hydrocarbons under the above-described approximations, the relevant basis set { $\chi$ } always is divisible into two *N*-dimensional subsets { $\chi^1$ } and { $\chi^2$ } so that pairs of orbitals belonging to chemical bonds (C=C or C–C) find themselves in different subsets [29]. This implies non-zero resonance parameters (1 and $\gamma$, respectively) to take place in the off-diagonal (intersubset) blocks of our initial Hamiltonian matrix ***H*** as exhibited below. Given that the AOs are enumerated in such a way that orbitals belonging to the same C=C bond (say, to the Ith one) acquire coupled numbers *i* and *N+i*, resonance parameters of these strong bonds take the diagonal positions in the intersubset blocks of the matrix ***H***.

Let us assume now that a certain number of heteroatoms (X) is introduced into the above-described parent polyene, the relevant electronegativities being slightly different from that of carbon atoms. Let the respective averaged alteration in the Coulomb parameter be denoted by $\alpha$ and supposed to be a first order term vs. the same energy unit. Obviously, a positive parameter $\alpha$ will correspond to the most common case of more electronegative heteroatoms.

As a result, the total Hamiltonian matrix ($H$) of our heteroatom(s)-containing compound is representable as a sum of a zero order member ($H_{(0)}$) and of the first order one ($H_{(1)}$). Moreover, both of these terms consist of four $N$x$N$-dimensional submatrices corresponding to subsets {$\chi^1$} and {$\chi^2$} and to their interaction, viz.

$$H = H_{(0)} + H_{(1)} = \begin{vmatrix} 0 & I \\ I & 0 \end{vmatrix} + \begin{vmatrix} \alpha A & \gamma B \\ \gamma B* & \alpha C \end{vmatrix}, \qquad (1)$$

where $I$ here and below stands for the unit (sub)matrix and the superscript $*$ designates the transposed (Hermitian-conjugate) matrix. The zero order member $H_{(0)}$ of Eq.(1) contains resonance parameters of C=C bonds, whilst the first order one $H_{(1)}$ embraces parameters of C–C bonds ($\gamma$) along with those of heteroatoms ($\alpha$). Since entire resonance parameters of C=C bonds are included into the zero order matrix $H_{(0)}$, diagonal elements of the submatrix $B$ take zero values (i.e. $B_{ii}=0$). Meanwhile, the relevant off-diagonal elements ($B_{ij}$, $i \neq j$) coincide with 1 in the positions associated with C–C bonds and vanish otherwise. By contrast, $A$ and $C$ are diagonal (sub)matrices containing unit elements in the positions referring to heteroatoms. Obviously, zero (sub)matrices $A$ and $C$ correspond to the parent polyene. It also deserves emphasizing that the particular constitution neither of the compound concerned nor of (sub)matrices $A$, $B$ and $C$ is specified here. Thus, the matrix $H$ of Eq.(1) actually represents the whole class of polyenes with heteroatoms.

The above-introduced subsets of AOs {$\chi^1$} and {$\chi^2$} are characterized by a zero intersubset energy gap. Consequently, perturbative approaches cannot be straightforwardly applied to the matrix $H$ of Eq.(1) and a certain transformation of the basis set is required. The simplest option here consists in turning to the basis of bond orbitals (BOs) of C=C bonds of the parent polyene. Let the bonding BO (BBO) of the Ith C=C bond $\phi_{(+)i}$ and its antibonding counterpart (ABO) $\phi_{(-)i}$ be defined as a normalized sum and difference, respectively, of the relevant AOs $\chi_i^1$ and $\chi_{N+i}^2$. Let the subsets of BBOs and of ABOs be correspondingly designated by {$\phi_{(+)}$} and {$\phi_{(-)}$}. Passing from the initial basis set {$\chi$} to the new one {$\phi$} is then representable by the following simple unitary matrix

$$U = \frac{1}{\sqrt{2}} \begin{vmatrix} I & I \\ I & -I \end{vmatrix}, \qquad (2)$$

which serves to transform the initial Hamiltonian matrix $H$ of Eq.(1). The new Hamiltonian matrix is then as follows

$$H' = H'_{(0)} + H'_{(1)} = \begin{vmatrix} I & 0 \\ 0 & -I \end{vmatrix} + \begin{vmatrix} T & R \\ R* & Q \end{vmatrix}, \qquad (3)$$

where the superscript ' serves to distinguish between the present matrix and that of Eq.(1). The $N$x$N$-dimensional submatrices $T$ and $Q$ now represent the interactions inside subsets {$\phi_{(+)}$} and {$\phi_{(-)}$}, respectively, whereas $R$ embraces intersubset interactions (i.e. resonance parameters between BBOs and ABOs). At the same time, all submatrices of the first order matrix $H'_{(1)}$ consist of zero and first order increments with respect to parameter $\alpha$ denoted below by superscripts (0) and (1), respectively. i.e.

$$T=T^{(0)}+T^{(1)}, \quad Q=Q^{(0)}+Q^{(1)}, \quad R=R^{(0)}+R^{(1)}, \qquad (4)$$

where

$$T^{(0)} = -Q^{(0)} = \frac{\gamma}{2}(B*+B), \quad R^{(0)} = -R^{(0)*} = \frac{\gamma}{2}(B*-B),$$

$$T^{(1)} = Q^{(1)} = \frac{\alpha}{2}(A+C), \quad R^{(1)} = R^{(1)*} = \frac{\alpha}{2}(A-C), \qquad (5)$$

Obviously, increments $T^{(0)}$, $Q^{(0)}$ and $R^{(0)}$ contain interactions between BOs of the parent polyene. Diagonal elements of these matrices vanish, i.e. $T_{ii}^{(0)} = Q_{ii}^{(0)} = R_{ii}^{(0)} = 0$ (The equality $B_{ii}=0$ should be recalled here). Meanwhile, off-diagonal elements of these matrices take non-zero values for BOs of first-neighbouring C=C bonds and vanish otherwise [29] (First-neighbouring formally-double bonds are defined as those connected by a formally-single bond). Further, the remaining terms of Eq.(4) (i.e. $T^{(1)}$, $Q^{(1)}$ and $R^{(1)}$) embrace corrections to the above-mentioned interactions due to introduction of heteroatom(s) and are diagonal matrices in contrast to the former. Non-zero elements $T_{ii}^{(1)}$ and $Q_{ii}^{(1)}$ (if any) represent newly-emerging self-interactions of BOs $\phi_{(+)i}$ and $\phi_{(-)i}$, respectively, whereas a significant element $R_{ii}^{(1)}$ indicates an additional resonance parameter to arise between the same BOs that is referred to below as the intrabond one. It also deserves mention that non-zero elements $R_{ii}^{(1)}$ actually refer to C=X bonds only. [Indeed, X=X bonds are characterized by uniform elements $A_{ii}=1$ and $C_{ii}=1$ and, consequently, the relevant elements $R_{ii}^{(1)}$ vanish]. Finally, all increments of Eq.(5) are symmetric matrices except for the only skew-symmetric component $R^{(0)}$.

Comparison of the matrix $H'$ of Eq.(3) to that underlying the PNCMO theory (see *Methods*) shows the former to coincide with a particular case of the latter. That is why the achievements of this theory are straightforwardly applicable to polyenes with heteroatom(s). The following points deserve mention here:

i) Occupation numbers (populations) of basis orbitals and total energies are expressible in this theory in terms of certain principal matrices $G_{(k)}$, $k=1,2,3..$ specified below and/or their derivatives $D_{(k)}^{(+)}$ and $D_{(k)}^{(-)}$, $k=2,3...$ defined as follows

$$D_{(2)}^{(+)} = G_{(1)}G_{(1)}^*, \qquad D_{(2)}^{(-)} = G_{(1)}^* G_{(1)}, \qquad D_{(3)}^{(+)} = G_{(1)}G_{(2)}^* + G_{(2)}G_{(1)}^*,$$
$$D_{(2)}^{(-)} = G_{(1)}^* G_{(2)} + G_{(2)}^* G_{(1)}, etc., \qquad (6)$$

where subscripts of all matrices indicate the order of the latter with respect to parameter(s) contained within the first order term of the Hamiltonian matrix. Meanwhile, superscripts of matrices $D_{(k)}^{(+)}$ and $D_{(k)}^{(-)}$ point to their links to subsets $\{\phi_{(+)}\}$ and $\{\phi_{(-)}\}$, respectively, as illustrated below.

ii) Matrices $G_{(k)}$ themselves are directly related to entire blocks (submatrices) of the relevant Hamiltonian matrix. For the particular form of the latter shown in Eq.(3), the above-mentioned relations take an algebraic form, viz.

$$G_{(1)} = -\frac{1}{2}R, \qquad G_{(2)} = \frac{1}{4}(TR - RQ), etc \qquad (7)$$

iii) Individual elements of matrices $G_{(k)}$, $D_{(k)}^{(+)}$ and $D_{(k)}^{(-)}$ have a clear chemical meaning, viz. these represent definite types of local and semi-local interactions of the relevant basis orbitals.

To illustrate the above-enumerated points, let us dwell on the Hamiltonian matrix $H'$ of Eq.(3). Let the occupation numbers of BOs $\phi_{(+)i}$ and $\phi_{(-)l}$ be denoted by $X_i^{(+)}$ and $X_l^{(-)}$, respectively. The expressions for the latter then take the form of the following power series

$$X_i^{(+)} = 2 + \sum_{k=2}^{\infty} X_{(k)ii}^{(+)}, \qquad X_l^{(-)} = \sum_{k=2}^{\infty} X_{(k)ll}^{(-)}, \qquad (8)$$

where 2 represents the initial (zero order) population of the BBO $\phi_{(+)i}$, whilst contributions of the $k$th order ($X_{(k)ii}^{(+)}$ and $X_{(k)ll}^{(-)}$) are proportional to respective diagonal elements of matrices of Eq.(6), viz.

$$X_{(k)ii}^{(+)} = -2D_{(k)ii}^{(+)}, \qquad X_{(k)ll}^{(-)} = 2D_{(k)ll}^{(-)}. \qquad (9)$$

Expressibility of matrices $D_{(k)}^{(+)}$ and $D_{(k)}^{(-)}$ via products of matrices $G_{(k)}$ of lower orders seen from Eq.(6), in turn, allows the contributions to occupation numbers to be represented as sums of partial increments of separate BOs of the opposite subset, viz.

$$X_{(k)ii}^{(+)} = -\sum_{(-)l} x_{(+)i,(-)l}^{(k)}, \qquad X_{(k)ll}^{(-)} = \sum_{(+)i} x_{(-)l,(+)i}^{(k)}, \qquad (10)$$

where $x_{(+)i,(-)l}^{(k)}$ may be exemplified as follows

$$x_{(+)i,(-)l}^{(2)} = 2(G_{(1)il})^2 > 0, \qquad x_{(+)i,(-)l}^{(3)} = 4G_{(1)il}G_{(2)il}, etc \qquad (11)$$

and describe partial populations of the $k$th order transferred inside the pair of orbitals $\phi_{(+)i}$ and $\phi_{(-)l}$ (Note that $x_{(+)i,(-)l}^{(k)}$ coincides with $x_{(-)l,(+)i}^{(k)}$ in accordance with the charge conservation condition). Sums of Eq.(10) over $(-)l$ and over $(+)i$ correspondingly embrace all ABOs and all BBOs of the given system. The minus sign of the first relation of Eq.(10) is introduced for convenience and expresses an aggreement that a positive partial population ($x_{(+)i,(-)l}^{(k)} > 0$) corresponds to the charge transfer from the BBO ($\phi_{(+)i}$) to the ABO ($\phi_{(-)l}$) that is accompanied by reduction of the total occupation number of the former and by an increase of that of the latter. The *a priori* positive sign of the second order increment $x_{(+)i,(-)l}^{(2)}$ (seen from Eq.(11)) also deserves attention here.

Total energies of polyenes with heteroatom(s) ($E$) also take the form of an analogous power series. Individual members of the latter $E_{(k)}$ ($k=0,1,2,…$) result from the general expression of the PNCMO theory (see *Methods*) after an additional employment of the first relation of Eq.(7). The starting members of the series concerned are as follows

$$E_{(0)} = 2N, \qquad E_{(1)} = 2TrT, \qquad E_{(2)} = 4Tr(G_{(1)}G_{(1)}^*) = 4TrD_{(2)}^{(+)},$$
$$E_{(3)} = 4Tr(G_{(2)}G_{(1)}^*) = 2TrD_{(3)}^{(+)}, \qquad (12)$$

where $Tr$ here and below stands for a trace of a matrix. Accordingly, the expression for $E_{(4)}$ contains the subsequent matrix $D_{(4)}^{(+)}$ depending on the third order matrix $G_{(3)}$ and so forth. This form of the correction $E_{(4)}$, however, proved to be rather cumbersome in practice and has been reformulated in terms of matrices of lower orders [33]. As a result, the fourth order energy ($E_{(4)}$) has been represented as a sum of a positive component ($E_{(4+)}$) and of a negative one ($E_{(4-)}$), i.e.

$$E_{(4)} = E_{(4+)} + E_{(4-)}, \qquad (13)$$

where

$$E_{(4+)} = 4Tr(G_{(2)}G_{(2)}^*), \qquad E_{(4-)} = -4Tr(D_{(2)}^{(+)})^2. \qquad (14)$$

It is seen that $E_{(4)}$ depends on all elements of matrices $G_{(2)}$ and $D_{(2)}^{(+)}$ in this case.

Let us now turn to interpretation of elements of our principal matrices and start with $G_{(1)}$ and $G_{(2)}$ defined by Eq.(7). The element $G_{(1)il}$ of the first order matrix $G_{(1)}$ is proportional to the resonance parameter ($R_{il}$) between BOs $\phi_{(+)i}$ and $\phi_{(-)l}$ and inversely proportional to the relevant energy gap (equal to 2). Thus, $G_{(1)il}$ represents the direct (through-space) interaction of above-mentioned BOs. Accordingly, a diagonal element $G_{(1)ii}$ describes an analogous interaction between BOs $\phi_{(+)i}$ and $\phi_{(-)i}$ inside the Ith formally-double bond. Furthermore, the element $G_{(2)il}$ of the second order matrix $G_{(2)}$ is interpretable as the indirect (through-bond) interaction of the same BOs via a single mediator. Indeed, the second relation of Eq.(7) yields the following expression for the particular element $G_{(2)il}$, viz.

$$G_{(2)il} = \frac{1}{4}[\sum_{(+)m} T_{im}R_{ml} - \sum_{(-)m} R_{im}Q_{ml}], \qquad (15)$$

where sums over (+)$m$ and over (−)$m$ embrace all BBOs and all ABOs as previously. It is seen that both BBOs $\phi_{(+)m}$ and ABOs $\phi_{(-)m}$ are able to play the role of mediators of this interaction. Additivity of contributions of individual mediators to the overall interaction $G_{(2)il}$ also deserves mention here.

More information about elements of the present matrices $G_{(1)}$ and $G_{(2)}$ follows after substituting Eq.(4) into Eq.(7). It is then seen that matrices $G_{(1)}$ and $G_{(2)}$ now contain terms of different orders with respect to parameter $\alpha$, viz.

$$G_{(1)} = G_{(1)}^{(0)} + G_{(1)}^{(1)}, \qquad G_{(2)} = G_{(2)}^{(0)} + G_{(2)}^{(1)} + G_{(2)}^{(2)}, \qquad (16)$$

where the zero order increments $G_{(1)}^{(0)}$ and $G_{(2)}^{(0)}$ correspondingly embrace direct and indirect interactions of BOs of the parent polyene studied previously [29, 30], whilst the remaining terms contain corrections to these interactions due to introduction of heteroatom(s) to be analyzed in the next subsection. That is why we confine ourselves here to a brief summary of results concerning matrices $G_{(1)}^{(0)}$ and $G_{(2)}^{(0)}$.

As with the matrix $R^{(0)}$ of Eq.(5), $G_{(1)}^{(0)}$ also is a skew-symmetric matrix and, consequently, contains zero elements in its diagonal positions, i.e. $G_{(1)ii}^{(0)} = 0$. At the same time, non-zero off-diagonal elements $(G_{(1)il}^{(0)} \neq 0, i \neq l)$ correspond only to BOs ($\phi_{(+)i}$ and $\phi_{(-)l}$) of first-neighbouring C=C bonds (I and L) and these significant elements are of uniform absolute values coinciding with $\gamma/4$. Furthermore, the expression for the second order matrix $G_{(2)}^{(0)}$ resembles that of Eq.(7) except for zero order matrices $T^{(0)}$, $Q^{(0)}$ and $R^{(0)}$ of Eq.(5) standing instead of the total ones (i.e. $T$, $Q$ and $R$, respectively). The skew-symmetric nature of the matrix $G_{(2)}^{(0)}$ then easily follows along with its zero diagonal elements ($G_{(2)ii}^{(0)} = 0$). Moreover, the mediating BO of the relevant indirect interaction (either a BBO $\phi_{(+)m}$ or an ABO $\phi_{(-)m}$) necessarily belongs to another (Mth) C=C bond that coincides neither with the Ith nor with the Lth one (The equality $T_{ii}^{(0)} = Q_{ii}^{(0)} = R_{ii}^{(0)} = 0$ should be recalled here). Finally, the mediating BO should overlap with both $\phi_{(+)i}$ and $\phi_{(-)l}$ to be an efficient mediator. These conditions are met for C=C bonds (M) taking place in between the Ith and the Lth one. As a result, non-zero indirect interactions ($G_{(2)il}^{(0)} \neq 0, i \neq l$) proved to be possible only for BOs $\phi_{(+)i}$ and $\phi_{(-)l}$ of second-neighbouring C=C bonds in polyenes [29].

Let us now dwell on the series of matrices $D_{(k)}^{(+)}$ and $D_{(k)}^{(-)}$ defined by Eq.(6) and starting with $k=2$. As opposed to their former analogues $G_{(k)}$, $k=2,3,...$, elements of the latter ($D_{(k)il}^{(+)}$ and $D_{(k)il}^{(-)}$) represent indirect interactions of orbitals inside the same subset, viz. between BBOs ($\phi_{(+)i}$ and $\phi_{(+)l}$) and between ABOs ($\phi_{(-)i}$ and $\phi_{(-)l}$), respectively. Thus, matrices $D_{(k)}^{(+)}$ and $D_{(k)}^{(-)}$ are correspondingly associated with subsets $\{\phi_{(+)}\}$ and $\{\phi_{(-)}\}$. It is also noteworthy that the above-specified indirect intrasubset interactions are mediated by a lower number of BOs as compared to intersubset ones underlying matrices $G_{(k)}$, $k=2,3,..$ For example, BOs of the opposite subset only are able to play this role in the second order interactions $D_{(2)il}^{(+)}$ and $D_{(2)il}^{(-)}$, viz. ABOs ($\phi_{(-)m}$) and BBOs ($\phi_{(+)m}$), respectively. Furthermore, $D_{(k)}^{(+)}$ and $D_{(k)}^{(-)}$ are symmetric matrices in contrast to $G_{(k)}$. This result easily follows from Eq.(6) after taking into account the skew-symmetric nature of matrices $G_{(k)}$, $k=1,2,…$ and implies non-zero diagonal elements $D_{(k)ii}^{(+)}$ and $D_{(k)ii}^{(-)}$ to be allowed. These elements accordingly represent indirect self-interactions of respective BOs. Finally, substituting Eq.(16) into Eq.(6) shows that expressions for matrices $\boldsymbol{D}_{(k)}^{(+)}$ and $\boldsymbol{D}_{(k)}^{(-)}$ also contain contributions of different orders with respect to parameter $\alpha$, viz.

$$\boldsymbol{D}_{(k)}^{(+)} = \sum_{s=0}^{k} \boldsymbol{D}_{(k)}^{(s+)} = \boldsymbol{D}_{(k)}^{(0+)} + \boldsymbol{D}_{(k)}^{(1+)} + ... + \boldsymbol{D}_{(k)}^{(k+)}, \quad \boldsymbol{D}_{(k)}^{(-)} = \sum_{s=0}^{k} \boldsymbol{D}_{(k)}^{(s-)} = \boldsymbol{D}_{(k)}^{(0-)} + \boldsymbol{D}_{(k)}^{(1-)} + ... + \boldsymbol{D}_{(k)}^{(k-)},$$
(17)

where increments of the same order ($s$) are interrelated as follows

$$\boldsymbol{D}_{(k)}^{(0+)} = \boldsymbol{D}_{(k)}^{(0-)}, \qquad \boldsymbol{D}_{(k)}^{(1+)} = -\boldsymbol{D}_{(k)}^{(1-)}, \qquad \boldsymbol{D}_{(k)}^{(2+)} = \boldsymbol{D}_{(k)}^{(2-)}, etc \qquad (18)$$

(The minus signs arise in these relations for terms of odd orders with respect to $\alpha$). The zero order contributions of Eq.(17) ($\boldsymbol{D}_{(k)}^{(0+)}$ and $\boldsymbol{D}_{(k)}^{(0-)}$) evidently refer to the parent polyene. The relevant diagonal elements ($D_{(k)ii}^{(0+)}$ and $D_{(k)ll}^{(0-)}$) determine the occupation numbers of BOs of C=C bonds ($\phi_{(+)i}$ and $\phi_{(-)l}$, respectively) as Eqs.(8) and (9) indicate. The first relation of Eq.(18) along with Eq.(9) then implies opposite signs and uniform absolute values of corrections of any order ($k$) to populations of BOs of the same C=C bond in polyenes (i.e the equality $X_{(k)ii}^{(0+)} = -X_{(k)ii}^{(0-)}$). Thus, no charge redistribution takes place among C=C bonds in the parent system in accordance with the expectation. In this context, a simple proportionality also deserves mention between any second order element $D_{(2)ii}^{(0+)}$ and the total number of first neighbours of the C=C bond concerned ($n_I$), viz. $D_{(2)ii}^{(0+)} = \gamma^2 n_I / 16$ [30]. Meanwhile, the third order corrections $D_{(3)ii}^{(0+)}$ take zero values in polyenes [30].

Now, let us turn to $\alpha$-dependent contributions of matrices $\boldsymbol{G}_{(k)}$, $\boldsymbol{D}_{(k)}^{(+)}$ and $\boldsymbol{D}_{(k)}^{(-)}$.

***Heteroatom influence upon local interactions of bond orbitals and their populations***

Let us start with the first order correction $\boldsymbol{G}_{(1)}^{(1)}$ of Eq.(16). As is seen from Eqs.(4), (5) and (7), the latter is proportional to the matrix $\boldsymbol{R}^{(1)}$ of Eq.(5). As a result, non-zero elements arise in the correction $\boldsymbol{G}_{(1)}^{(1)}$ only in its diagonal positions referring to C=X bonds. Moreover, absolute values of these elements are uniform and equal to $\alpha/4$. Thus, introduction of a heteroatom into a certain (Ith) C=C bond of the parent polyene gives birth to a local direct interaction between the BBO and the ABO of this bond ($\phi_{(+)i}$ and $\phi_{(-)i}$). This result causes little surprise if we bear in mind that BOs of C=C bonds of the parent polyene $\{\phi\}$ defined by Eq.(2) are no longer eigenfunctions of the total Hamiltonian matrix block embracing a C=X bond.

Let us now turn to corrections $\boldsymbol{G}_{(2)}^{(1)}$ and $\boldsymbol{G}_{(2)}^{(2)}$ of Eq.(16) that are expressible via products of matrices of Eq.(5). In particular, the definition of the correction $\boldsymbol{G}_{(2)}^{(2)}$ closely resembles that of the total matrix $\boldsymbol{G}_{(2)}$ itself shown in Eq.(7) except for matrices $\boldsymbol{T}^{(1)}, \boldsymbol{Q}^{(1)}$ and $\boldsymbol{R}^{(1)}$ standing for $\boldsymbol{T}, \boldsymbol{Q}$ and $\boldsymbol{R}$, respectively. The diagonal constitution of the former along with coincidence between matrices $\boldsymbol{T}^{(1)}$ and $\boldsymbol{Q}^{(1)}$ seen from Eq.(5) then yields the equality $\boldsymbol{G}_{(2)}^{(2)} = 0$. Meanwhile, the definition of the remaining correction $\boldsymbol{G}_{(2)}^{(1)}$ contains all (i.e. six) matrices of Eq.(5). Moreover, $\boldsymbol{G}_{(2)}^{(1)}$ is a symmetric matrix in contrast to the zero order increment $\boldsymbol{G}_{(2)}^{(0)}$ of Eq.(16). An element $G_{(2)il}^{(1)}$ of this matrix is expressible as follows

$$G_{(2)il}^{(1)} = \frac{1}{4}(T_{ii}^{(1)} R_{il}^{(0)} - R_{ii}^{(1)} Q_{il}^{(0)} + T_{il}^{(0)} R_{ll}^{(1)} - R_{il}^{(0)} Q_{ll}^{(1)}) \qquad (19)$$

and depends on elements of matrices of Eq.(5) referring to the Ith and the Lth double bonds only. At the same time, the correction $G_{(2)il}^{(1)}$ itself describes a certain additional indirect interaction between BOs of the same double bonds I and L ($\phi_{(+)i}$ and $\phi_{(-)l}$). Separate terms of the right-hand side of Eq.(19) correspondingly represent increments of four potential mediators of this interaction, viz. of BOs $\phi_{(+)i}$, $\phi_{(-)i}$, $\phi_{(+)l}$ and $\phi_{(-)l}$. Thus, the interacting orbitals $\phi_{(+)i}$ and $\phi_{(-)l}$ also are among potential mediators here, i.e. self-mediating effects of the latter are possible. The first order magnitude of the indirect interactions $G^{(1)}_{(2)il}$ with respect to parameter $\alpha$ also is easily seen from Eq.(19). Finally, a non-zero value condition for corrections $G^{(1)}_{(2)il}$ may be formulated on the basis of Eq.(19): First, presence of at least a single heteroatom within either of double bonds concerned (i.e. either within the Ith or within the Lth one) is among necessary conditions (Recall that significant diagonal elements of matrices $\boldsymbol{T}^{(1)}$, $\boldsymbol{Q}^{(1)}$ and $\boldsymbol{R}^{(1)}$ are entirely due to heteroatoms). Second, a first neighbourhood of the Ith and Lth double bonds is imperative (As already mentioned, non-zero elements $T^{(0)}_{il}$, $Q^{(0)}_{il}$ and $R^{(0)}_{il}$ refer only to BOs of first-neighbouring C=C bonds in polyenes). In summary, non-zero corrections $G^{(1)}_{(2)il}$ are possible only between the BBO $\phi_{(+)i}$ of a heteroatom(s)-containing double bond (I) and the ABO $\phi_{(-)l}$ of its first neighbour (L) of any constitution (or vice versa), i.e. for two-membered conjugated fragments (I–L) containing at least a single heteroatom (X). [Note that indirect interactions between BOs of the same double bond (e.g. $G^{(1)}_{(2)ii}$) vanish because of zero diagonal elements of matrices $\boldsymbol{T}^{(0)}$, $\boldsymbol{Q}^{(0)}$ and $\boldsymbol{R}^{(0)}$]. In addition, the above-discussed local nature of elements $G^{(1)}_{(2)il}$ allows us to expect the relevant indirect interactions to be transferable for the same fragment I–L inside different compounds. [It deserves mention here, however, that the above-formulated condition is not sufficient to ensure a non-zero element $G^{(1)}_{(2)il}$. The point is that cancellation of terms of the right-hand side of Eq.(19) is possible as demonstrated below]. Let us now return to the second relation of Eq.(16) and note that the zero order term $G^{(0)}_{(2)il}$ (the above-discussed correction $G^{(1)}_{(2)il}$ is added to) vanishes for first-neighbouring double bonds I and L in parent polyenes [29, 30]. Thus, a non-zero correction $G^{(1)}_{(2)il}$ (if any) implies emergence of a new indirect interaction between BOs of first-neighbouring double bonds I and L after introduction of heteroatom(s). It is also seen that local (direct) interactions of BOs inside a heteroatom-containing double bond (e. g. self-interactions of BOs inside the Ith bond ($T^{(1)}_{ii}$ and $Q^{(1)}_{ii}$) along with the relevant intrabond resonance parameter $R^{(1)}_{ii}$) potentially give birth to semi-local (indirect) interactions $G^{(1)}_{(2)il}$ embracing two-membered conjugated fragments I–L, the double bond concerned (i.e. the Ith one) participates in.

Let us now turn to $\alpha$-dependent increments of matrices $\boldsymbol{D}^{(+)}_{(k)}$ and $\boldsymbol{D}^{(-)}_{(k)}$ (see Eq.(17)). Due to simple relations of Eq.(18), we can confine ourselves to analysis of increments of matrices $\boldsymbol{D}^{(+)}_{(k)}$. Let us start with corrections $\boldsymbol{D}^{(1+)}_{(2)}$ and $\boldsymbol{D}^{(2+)}_{(2)}$ to the second order matrix $\boldsymbol{D}^{(0+)}_{(2)}$. The increment $\boldsymbol{D}^{(1+)}_{(2)}$ is the intrasubset analogue of the above-discussed correction $\boldsymbol{G}^{(1)}_{(2)}$. It is then no surprise that an element of this matrix ($D^{(1+)}_{(2)il}$) also depends on those of matrices of Eq.(5) pertinent to the Ith and Lth double bonds, viz.

$$D^{(1+)}_{(2)il} = \frac{1}{4}[-R^{(1)}_{ii}R^{(0)}_{il} + R^{(0)}_{il}R^{(1)}_{ll}]. \quad (20)$$

Analysis of Eq.(20) also closely resembles that of Eq.(19) except for a lower number of potential mediators of the underlying indirect interaction. Indeed, two BOs only (viz. the ABOs $\phi_{(-)i}$ and $\phi_{(-)l}$) are now able to play this role instead of four BOs in the case of Eq.(19) (Note that the two terms of the right-hand side of Eq.(20) represent just the increments of ABOs $\phi_{(-)i}$ and $\phi_{(-)l}$, respectively). In spite of the above-mentioned difference, however, non-zero off-diagonal elements $D_{(2)il}^{(1+)}, i \neq l$ also refer to two-membered conjugated fragments (I–L) with at least a single heteroatom (X), whilst the relevant diagonal elements $D_{(2)ii}^{(1+)}$ take zero values. The latter circumstance, in turn, implies no consequences of the correction $D_{(2)}^{(1+)}$ upon populations of BOs (see Eq.(9)).

The remaining correction to the same second order matrix $D_{(2)}^{(0+)}$ (i.e. $D_{(2)}^{(2+)}$) is expressible as square of the above-considered matrix $G_{(1)}^{(1)}$. As with the latter, the matrix $D_{(2)}^{(2+)}$ then also contains non-zero elements of uniform absolute values in the diagonal positions associated with C=X bonds. Due to relations of Eq.(18), the same consequently refers to its counterpart $D_{(2)}^{(2-)}$. We then obtain that

$$D_{(2)ii}^{(2+)} = D_{(2)ii}^{(2-)} = (G_{(1)ii}^{(1)})^2 = \frac{\alpha^2}{16} > 0 \qquad (21)$$

for any (Ith) C=X bond whatever the total number of the latter. Employment of Eqs.(9)-(11), in turn, yields the relevant alterations in occupation numbers of BOs $\phi_{(+)i}$ and $\phi_{(-)i}$, viz.

$$X_{(2)ii}^{(2+)loc} = -\frac{\alpha^2}{8} < 0, \qquad X_{(2)ii}^{(2-)loc} = \frac{\alpha^2}{8} > 0, \qquad (22)$$

as well as the additional partial population ($x_{(+)i,(-)i}^{(2)loc}$) transferred from the former orbital to the latter coinciding with $\alpha^2/8$. The local nature of this charge redistribution is evident and this fact is indicated here and below by superscripts *loc*. As this redistribution actually takes place inside separate C=X bonds, it is also conveniently referred to as the intrabond one.

As is seen from the above discussion, corrections both $D_{(2)}^{(1+)}$ and $D_{(2)}^{(2+)}$ make no contributions to charge redistribution among formally-double bonds (i.e. to the interbond one). That is why $\alpha$-dependent increments of the third order matrix $D_{(3)}^{(+)}$ deserve our attention. Due to the above-established equality $G_{(2)}^{(2)}=0$, the increment of the highest order ($s=3$) with respect to parameter $\alpha$ (i.e. $D_{(3)}^{(3+)}$) vanishes in Eq.(17). Meanwhile, contributions $D_{(3)}^{(1+)}$ and $D_{(3)}^{(2+)}$ exhibit "an opposite behaviour" as compared to their above-considered second order analogues $D_{(2)}^{(1+)}$ and $D_{(2)}^{(2+)}$: The matrix $D_{(3)}^{(2+)}$ contains zero diagonal elements due to the equality $G_{(2)ii}^{(1)} = 0$ and makes no contribution to the interbond charge transfer, whereas analogous elements of the first order term with respect to $\alpha$ ($D_{(3)ii}^{(1+)}$) take significant values. Employment of Eqs. (9) and (18) then consequently results into following expressions for the relevant alterations in occupation numbers of BOs of the Ith double bond $\phi_{(+)i}$ and $\phi_{(-)i}$, viz.

$$X_{(3)ii}^{(1+)} = -4\sum_{(-)l} G_{(1)il}^{(0)} G_{(2)il}^{(1)}, \qquad X_{(3)ii}^{(1-)} = 4\sum_{(+)l} G_{(1)li}^{(0)} G_{(2)li}^{(1)}. \qquad (23)$$

The equality $X_{(3)ii}^{(1+)} = X_{(3)ii}^{(1-)}$ easily results from Eq.(23) (after taking into account the relations $G_{(1)il}^{(0)} = -G_{(1)li}^{(0)}$ and $G_{(2)il}^{(1)} = G_{(2)li}^{(1)}$) and implies that the BOs of the Ith both $\phi_{(+)i}$ and $\phi_{(-)i}$ both either loose or acquire an additional population due to introduction of heteroatom(s). Further, expressions of Eq.(23) formally contain sums over BOs of all double bonds (L) of the given system. Actually, however, these sums embrace BOs of two-membered conjugated fragments (I–L) with at least a single heteroatom (X), the Ith double bond participates in. [To show this, no more is required as to recall the above-established constitution of matrices $G_{(2)}^{(1)}$]. A subsequent employment of Eqs.(10) and (11) then yields the following expressions for additional partial populations transferred inside an individual fragment I–L, viz.

$$x^{(3)}_{(+)i,(-)l} = 4G^{(0)}_{(1)il}G^{(1)}_{(2)il}, \qquad x^{(3)}_{(-)i,(+)l} = 4G^{(0)}_{(1)li}G^{(1)}_{(2)li} \qquad (24)$$

that, in turn, are interrelated as follows

$$x^{(3)}_{(+)i,(-)l} = x^{(3)}_{(-)l,(+)i} = -x^{(3)}_{(-)i,(+)l} = -x^{(3)}_{(+)l,(-)i}. \qquad (25)$$

Thus, the additional population transferred between orbitals $\phi_{(+)i}$ and $\phi_{(-)l}$ and its counterpart referring to the remaining BOs $\phi_{(+)l}$ and $\phi_{(-)i}$ are of opposite signs and of uniform absolute values. For example, if the BBO $\phi_{(+)i}$ acquires a certain population from the ABO $\phi_{(-)l}$ (i.e. $x^{(3)}_{(+)i,(-)l} < 0$), an analogous population is transferred from the BBO $\phi_{(+)l}$ to the ABO $\phi_{(-)i}$ (i.e. $x^{(3)}_{(+)l,(-)i} > 0$) or vicee versa. In other words, population is predicted to be unambiguosly transferred either from the Ith bond to the Lth one or vice versa inside a conjugated fragment I–L. Moreover, the ultimate direction of this charge redistribution depends on signs of elements $G^{(0)}_{(1)il}$ and $G^{(1)}_{(2)il}$: Given that these signs are uniform (opposite), the Ith double bond looses (acquires) an additional population and plays the role of the electron-donating (accepting) bond inside the given fragment I–L. Obviously, no charge redistribution is expected for fragments characterized by zero corrections $G^{(1)}_{(2)il}$.

Let us now return to total occupation numbers of BOs of our heteroatom(s)-containing system defined by Eq.(8). As is seen after an additional employment of Eqs.(9), (17) and (21)-(23), the occupation numbers concerned also consist of sums of contributions of different orders ($s$) with respect to parameter $\alpha$, viz. (i) of the zero order ones referring to the parent polyene, (ii) of the first order corrections of Eq.(23) embracing non-local charge redistributions among formally-double bonds and consisting of sums of transferable increments of separate conjugated fragments I–L and (iii) of the second order corrections of Eq.(22) representing local charge redistributions inside individual formally-double bonds due to passing from C=C bonds to C=X ones.

*Analysis of conjugation energies*

Let us now turn to the expansion for total energies ($E$) of heteroatom(s)-containing polyenes shown in Eqs.(12)-(14). The zero order member of this series ($E_{(0)}$) coincides with the sum of energies of $N$ isolated C=C bonds of the parent polyene ($E^{(0)}_{(0)}$). By contrast, a non-zero first order correction (if any) is entirely associated with heteroatom(s) and is proportional to the parameter $\alpha$. Indeed, employment of Eqs.(4) and (5) within the second relation of Eq.(12) yields the following expression

$$E_{(1)} = E^{(1)loc}_{(1)} = 2Tr\boldsymbol{T}^{(1)} = \alpha N_{C=X} + 2\alpha N_{X=X} > 0, \qquad (26)$$

where zero diagonal elements of the matrix $\boldsymbol{T}^{(0)}$ are taken into consideration. Notations $N_{C=X}$ and $N_{X=X}$ stand here for total numbers of C=X and X=X bonds, respectively. Thus, the correction $E_{(1)}$ actually consists of local increments of separate C=X and/or X=X bonds, where the contribution of an individual bond represents the alteration in the energy of the latter due to introduction of heteroatom(s). The positive sign of the correction $E_{(1)}$, in turn, implies an increase of the overall stability of the system after substitution. It is also seen that the correction $E_{(1)}$ is proportional to the total number of heteroatoms present in the given system. This result coincides with those of the CLHPT [22, 23] and of the chemical graph theory [18, 19, 21] and causes little surprise.

Further, the second order correction of the same expansion ($E_{(2)}$) depends on diagonal elements of the matrix $\boldsymbol{D}^{(+)}_{(2)}$ as the respective relation of Eq.(12) shows. After employment of

Eqs.(17) and (21) along with the above-established zero diagonal elements of the matrix $\mathbf{D}_{(2)}^{(1+)}$, we obtain that

$$E_{(2)} = E_{(2)}^{(0)} + E_{(2)}^{(2)loc} = \frac{\gamma^2}{2}(N-1) + \frac{\alpha^2}{4}N_{C=X} > 0. \tag{27}$$

Thus, the correction $E_{(2)}$ consists of a sum of zero and second order components with respect to $\alpha$ and also is a positive quantity. The component $E_{(2)}^{(0)}$ represents the second order energy of the parent polyene and is proportional to the relevant number of formally-single (C–C) bonds and/or butadiene-like fragments ($N$–1) [29, 30]. Meanwhile, the remaining component ($E_{(2)}^{(2)loc}$) originates from the matrix $\mathbf{D}_{(2)}^{(2+)}$ (see Eq.(21)) and contains a sum of local contributions of individual C=X bonds. Besides, the increment of a single C=X bond ($\alpha^2/4$) coincides with the alteration in the $\pi$-electron energy of ethene after introduction of a heteroatom following from the CLHPT. (Note that the self-polarizability of an atom of ethene equals to 1/2). Analogously, the third order energy of our heteroatom(s)-containing polyene ($E_{(3)}$) is determined by diagonal elements of the matrix $\mathbf{D}_{(3)}^{(+)}$ and formally contains terms up to the third order with respect to parameter $\alpha$ (s=0,1,2,3). If we recall, however, the zero value of third order energy of any polyene ($E_{(3)}^{(0)} = 0$) established previously [29, 30] along with the above-derived equalities $D_{(3)ii}^{(3+)} = D_{(3)ii}^{(2+)} = 0$, we obtain that

$$E_{(3)} = E_{(3)}^{(1+)} = 2Tr\mathbf{D}_{(3)}^{(1+)} = 4\sum_{(+))i(-)l}\sum G_{(1)il}^{(0)}G_{(2)il}^{(1)} = 0. \tag{28}$$

The zero value of the correction $E_{(3)}$ seen from Eq.(28) is due to different natures of matrices $\mathbf{G}_{(2)}^{(1)}$ and $\mathbf{G}_{(1)}^{(0)}$ (these are symmetric and skew-symmetric, respectively) and to a consequent cancellation of increments of pairs of BOs $\phi_{(+)i}$, $\phi_{(-)l}$ and $\phi_{(+)l}$, $\phi_{(-)i}$ in the sum of the right-hand side. Again, the above-mentioned increments are proportional to partial transferred populations $x_{(+)i,(-)l}^{(3)}$ and $x_{(+)l,(-)i}^{(3)}$ of Eq.(25). Thus, contributions to $E_{(3)}$ associated just with these populations cancel out one another here.

As is seen from the above discussion, contributions $E_{(k)}$ to within the third order ($k$=3) inclusive embrace the dependence of the total energy of our heteroatom(s)-containing system ($E$) only upon total numbers of double and single bonds of different types. To reveal the role of the mutual arrangement of these bonds, corrections of higher orders are then required and the fourth order energy $E_{(4)}$ of Eqs.(13) and (14) deserves attention in the first place. An analogous analysis of these expressions shows that terms of zero, first and second orders with respect to $\alpha$ are present inside the positive component $E_{(4+)}$, viz. $E_{(4+)}^{(0)}$, $E_{(4+)}^{(1)}$ and $E_{(4+)}^{(2)}$. Meanwhile, the negative component $E_{(4-)}$ embraces terms up to fourth order ($E_{(4-)}^{(0)}$, $E_{(4-)}^{(1)}$, $E_{(4-)}^{(2)}$, $E_{(4-)}^{(3)}$ and $E_{(4-)}^{(4)loc}$), the last one (i.e. $E_{(4-)}^{(4)loc}$) containing a sum of local fourth order corrections to energies of isolated C=X bonds and coinciding with $-\alpha^4 N_{C=X}/64$. [The correction $E_{(4-)}^{(4)loc}$ originates from the diagonal matrix $\mathbf{D}_{(2)}^{(2+)}$, an element of which is shown in Eq.(21)].

On the whole, the above-enumerated increments $E_{(k)}$ (including those of Eqs.(26)-(28)) may be collected to make a sum of three principal contributions to the total energy ($E$) of our heteroatom(s)-containing system, viz.

$$E = E^{(0)} + E^{loc} + \delta E_{(4)}, \tag{29}$$

where $E^{(0)}$ embraces all members of the expansion for the total energy of the parent polyene, i.e.

$$E^{(0)} = E^{(0)}_{(0)} + E^{(0)}_{(2)} + E^{(0)}_{(4)} + ... , \qquad (30)$$

the increment $E^{loc}$ consists of local corrections to energies of isolated double bonds due to introduction of heteroatoms (i.e. $E^{(1)loc}_{(1)}$, $E^{(2)loc}_{(2)}$ and $E^{(4)loc}_{(4-)}$) and $\delta E_{(4)}$ stands for the sum of the remaining "mixed" terms containing products of both parameters $\alpha$ and $\gamma$ (i.e $\alpha\gamma^3$, $\alpha^2\gamma^2$ and $\alpha^3\gamma$) and representing the non-local part of the fourth order energy $E_{(4)}$. The contribution $\delta E_{(4)}$, in turn, is expressible as a sum of the positive component ($\delta E_{(4+)}$) and of the negative one ($\delta E_{(4-)}$), where

$$\delta E_{(4+)} = E^{(1)}_{(4+)} + E^{(2)}_{(4+)}, \qquad \delta E_{(4-)} = E^{(1)}_{(4-)} + E^{(2)}_{(4-)} + E^{(3)}_{(4-)}. \qquad (31)$$

To reveal the meaning of the increment $\delta E_{(4)}$, let us recall the definition of the conjugation energy (CE) of any polyene [30] as the difference between the relevant total energy and the sum of energies of $N$ isolated C=C bonds. On the basis of Eq.(30), the CE of our parent polyene ($\Delta E^{(0)}$) is expressible as follows

$$\Delta E^{(0)} = E^{(0)} - E^{(0)}_{(0)} = E^{(0)}_{(2)} + E^{(0)}_{(4)} + ... \qquad (32)$$

The CE of the heteroatom(s)-containing compound ($\Delta E$) also may be defined analogously, where $E^{(0)}_{(0)} + E^{loc}$ plays the role of the sum of energies of isolated formally-double bonds. From Eq.(29), we then obtain that

$$\Delta E = E - (E^{(0)}_{(0)} + E^{loc}) = \Delta E^{(0)} + \delta E_{(4)}, \qquad (33)$$

where $\Delta E^{(0)}$ is shown in Eq.(32). As is seen from Eq.(33), the correction $\delta E_{(4)}$ represents the alteration in the CE when passing from the parent polyene to its heteroatom(s)-containing derivative. Let us now consider separate contributions to this alteration shown in Eq.(31).

Members of the positive component $\delta E_{(4+)}$ are expressible as follows

$$E^{(1)}_{(4+)} = 8\sum_{(+)i}\sum_{(-)l} G^{(1)}_{(2)il} G^{(0)}_{(2)il}, \qquad E^{(2)}_{(4+)} = 4\sum_{(+)i}\sum_{(-)l} (G^{(1)}_{(2)il})^2. \qquad (34)$$

The left increment of Eq.(34) ($E^{(1)}_{(4+)}$) vanishes due to different positions of non-zero elements of matrices $G^{(1)}_{(2)}$ and $G^{(0)}_{(2)}$ (As already mentioned, non-zero elements refer to BOs of first- and second-neighbouring double bonds, respectively, in these matrices). As a result, the overall positive component $\delta E_{(4+)}$ actually coincides with the remaining increment $E^{(2)}_{(4+)}$. The latter, in turn, consists of a sum of local and transferable contributions of individual two-membered conjugated fragments (I–L) containing at least a single heteroatom [The above-established constitution of the matrix $G^{(1)}_{(2)}$ should be recalled here]. Let the contribution of a particular fragment I–L to $E^{(2)}_{(4+)}$ (and/or to $\delta E_{(4+)}$) be denoted by $E^{(2)}_{(4+)I-L}$. If we observe in addition that two elements of the matrix $G^{(1)}_{(2)}$ (viz. $G^{(1)}_{(2)il}$ and $G^{(1)}_{(2)li}$) correspond to a single fragment I–L, we obtain that

$$E^{(2)}_{(4+)I-L} = 8(G^{(1)}_{(2)il})^2 > 0, \qquad (35)$$

i.e the contribution concerned is stabilizing for all fragments I–L described by a non-zero correction $G^{(1)}_{(2)il}$ whatever the sign of the latter.

Let us now turn to the component $\delta E_{(4-)}$ of Eq.(31). The increment $E^{(1)}_{(4-)}$ is an analogue of $E^{(1)}_{(4+)}$ and vanishes for similar reasons (viz. due to different positions of non-zero elements of underlying matrices $D^{(0+)}_{(2)}$ and $D^{(1+)}_{(2)}$). The same refers also to the increment $E^{(3)}_{(4-)}$ correspondingly depending on matrices $D^{(1+)}_{(2)}$ and $D^{(2+)}_{(2)}$. Thus, the whole component $\delta E_{(4-)}$ coincides with the remaining increment $E^{(2)}_{(4-)}$. As opposed to its former counterpart $E^{(2)}_{(4+)}$, however, the latter contains a sum of two subcomponents $E^{(2)a}_{(4-)}$ and $E^{(2)b}_{(4-)}$ representable as follows

$$E^{(2)a}_{(4-)} = -4\sum_{(+)i}\sum_{(-)l}(D^{(1+)}_{(2)il})^2, \qquad E^{(2)b}_{(4-)} = -8\sum_{(+)i}\sum_{(-)l}D^{(2+)}_{(2)il}D^{(0+)}_{(2)il}. \qquad (36)$$

The subcomponent $E^{(2)a}_{(4-)}$ actually is a complete analogue of the former component $E^{(2)}_{(4+)}$ and, consequently, it is also expressible as a sum of local and transferable contributions of two-membered heteroatom(s)-containing fragments I–L as exhibited below. Meanwhile, the subcomponent $E^{(2)b}_{(4-)}$ of Eq.(36) has no analogue among members of $\delta E_{(4+)}$ and arises due to the non-zero matrix $D^{(2+)}_{(2)}$ (in contrast to the zero matrix $G^{(2)}_{(2)}$). As already mentioned, $D^{(2+)}_{(2)}$ is a diagonal matrix containing elements $\alpha^2/16$ in the positions associated with C=X bonds only (see Eq.(21)). If we recall in addition the above-cited proportionality between an element of the matrix $D^{(0+)}_{(2)}$ (i.e. $D^{(0+)}_{(2)ii}$) and the total number of first neighbours of the Ith C=C bond in polyenes ($n_I$) [30], we obtain that

$$E^{(2)b}_{(4-)} = -\frac{\alpha^2\gamma^2}{32}\sum_{I(C=X)}n_I < 0, \qquad (37)$$

where the sum over I embraces C=X bonds. Nevertheless, the subcomponent $E^{(2)b}_{(4-)}$ also is representable as a sum of local and transferable increments of heteroatom(s)-containing conjugated fragments. Moreover, the contribution of a particular fragment I–L is proportional to the number of C=X bonds present inside the latter coinciding with either 0, 1 or 2 and denoted below by $n_{C=X}$. Indeed, the Lth first neighbour of the Ith C=X bond makes a contribution to $E^{(2)b}_{(4-)}$ eqal to $-\alpha^2\gamma^2/32$, if L stands for either C=C or X=X bond and $n_{C=X}$=1. Meanwhile, the analogous contribution takes a twofold value ($-\alpha^2\gamma^2/16$) if both I and L coincide with C=X bonds and $n_{C=X}$=2. Thus, the whole increment of the fragment I–L to $E^{(2)}_{(4-)}$ and/or to $\delta E_{(4-)}$ ($E^{(2)}_{(4-)I-L}$) is as follows

$$E^{(2)}_{(4-)I-L} = -8(D^{(1+)}_{(2)il})^2 - \frac{\alpha^2\gamma^2}{32}n_{C=X} < 0. \qquad (38)$$

It is seen, therefore, that the overall alteration in the CE due to substitution ($\delta E_{(4)}$) is expressible as a sum over all heteroatom(s)-containing two-membered conjugated fragments, viz.

$$\delta E_{(4)} = \sum_{I-L}\delta E_{(4)I-L}, \qquad (39)$$

where the contribution of an individual fragment $\delta E_{(4)I-L}$ consists of two components of opposite signs $E^{(2)}_{(4+)I-L}$ and $E^{(2)}_{(4-)I-L}$ shown in Eqs.(35) and (38), respectively, and correspondingly determined by squares of newly-emerging indirect inter- and intra-subset interactions $G^{(1)}_{(2)il}$ and $D^{(1+)}_{(2)il}$. Thus, the actual sign and the absolute value of the overall contribution $\delta E_{(4)I-L}$ is expected to depend upon the ratio between absolute values of these additional interactions that, in turn, is likely to be determined by the constitution of the given fragment. The next subsection addresses just this point.

*Discussion of specific fragments*

Nine different two-membered conjugated fragments (I–L) are possible in polyenes with heteroatom(s) that are exhibited in the left part of Table 1. Obviously, fragments I-IX

Table 1

Characteristics of conjugated fragments (I-IX) consisting of two connected formally-double bonds I and L: Numbers of C=X bonds ($n_{C=X}$), additional indirect interactions between the BBO $\phi_{(+)i}$ and the ABO $\phi_{(-)l}$ ($G^{(1)}_{(2)il}$), analogous interactions between BBOs $\phi_{(+)i}$ and $\phi_{(+)l}$ ($D^{(1+)}_{(2)il}$), partial populations transferred from the BBO $\phi_{(+)l}$ to the ABO $\phi_{(-)i}$ ($x^{(3)}_{(+)l,(-)i}$) and contributions of individual fragments to the overall alteration in the CE ($\delta E_{(4)I-L}$) along with their positive and negative components ($E^{(2)}_{(4+)I-L}$ and $E^{(2)}_{(4-)I-L}$, respectively). Classification of fragments into two groups (A and B) is shown in the left column

| Nr. | Structure | $n_{C=X}$ | $G^{(1)}_{(2)il}$ | $D^{(1+)}_{(2)il}$ | $x^{(3)}_{(+)l,(-)i}$ | $E^{(2)}_{(4+)I-L}$ | $E^{(2)}_{(4-)I-L}$ | $\delta E_{(4)I-L}$ | |
|---|---|---|---|---|---|---|---|---|---|
| I | X = C – C = C | 1 | $\dfrac{\alpha\gamma}{8}$ | $-\dfrac{\alpha\gamma}{16}$ | $\dfrac{\alpha\gamma^2}{8}$ | $\dfrac{\alpha^2\gamma^2}{8}$ | $-\dfrac{\alpha^2\gamma^2}{16}$ | $\dfrac{\alpha^2\gamma^2}{16}$ | A |
| II | C = X – C = C | 1 | 0 | $\dfrac{\alpha\gamma}{16}$ | 0 | 0 | $-\dfrac{\alpha^2\gamma^2}{16}$ | $-\dfrac{\alpha^2\gamma^2}{16}$ | B |
| III | X = C – X = C | 2 | $\dfrac{\alpha\gamma}{8}$ | 0 | $\dfrac{\alpha\gamma^2}{8}$ | $\dfrac{\alpha^2\gamma^2}{8}$ | $-\dfrac{\alpha^2\gamma^2}{16}$ | $\dfrac{\alpha^2\gamma^2}{16}$ | A |
| IV | X = C – C = X | 2 | 0 | $-\dfrac{\alpha\gamma}{8}$ | 0 | 0 | $-\dfrac{3\alpha^2\gamma^2}{16}$ | $-\dfrac{3\alpha^2\gamma^2}{16}$ | B |
| V | C = X – X = C | 2 | 0 | $\dfrac{\alpha\gamma}{8}$ | 0 | 0 | $-\dfrac{3\alpha^2\gamma^2}{16}$ | $-\dfrac{3\alpha^2\gamma^2}{16}$ | B |
| VI | X = X – C = C | 0 | $\dfrac{\alpha\gamma}{8}$ | 0 | $\dfrac{\alpha\gamma^2}{8}$ | $\dfrac{\alpha^2\gamma^2}{8}$ | 0 | $\dfrac{\alpha^2\gamma^2}{8}$ | A |
| VII | X = X – X = C | 1 | $\dfrac{\alpha\gamma}{8}$ | $\dfrac{\alpha\gamma}{16}$ | $\dfrac{\alpha\gamma^2}{8}$ | $\dfrac{\alpha^2\gamma^2}{8}$ | $-\dfrac{\alpha^2\gamma^2}{16}$ | $\dfrac{\alpha^2\gamma^2}{16}$ | A |
| VIII | X = X – C = X | 1 | 0 | $-\dfrac{\alpha\gamma}{16}$ | 0 | 0 | $-\dfrac{\alpha^2\gamma^2}{16}$ | $-\dfrac{\alpha^2\gamma^2}{16}$ | B |
| IX | X = X – X = X | 0 | 0 | 0 | 0 | 0 | 0 | 0 | B |

coincide with derivatives of butadiene containing different numbers of heteroatoms (X) in various positions. Let the four carbon atoms of the parent butadiene be enumerated as follows $C^1_i = C^2_{N+i} - C^1_l = C^2_{N+l}$, where the 2p$_z$ AOs of carbon atoms under numbers $i, l$ and $N+i, N+l$ are supposed to belong to subsets $\{\chi^1\}$ and $\{\chi^2\}$, respectively, as indicated by superscripts. The same numbering of AOs and/or of atoms will be preserved also when passing to fragments I-IX. As a result, elements of $\alpha$-independent matrices of Eqs.(4), (5), (7) and (16) are as follows

$$T^{(0)}_{il} = -Q^{(0)}_{il} = R^{(0)}_{il} = -R^{(0)}_{li} = \frac{\gamma}{2}, \qquad G^{(0)}_{(1)il} = -G^{(0)}_{(1)li} = -\frac{\gamma}{4} \qquad (40)$$

for all fragments concerned. Meanwhile, elements of $\alpha$-dependent matrices $\mathbf{T}^{(1)}$, $\mathbf{Q}^{(1)}$ and $\mathbf{R}^{(1)}$ depend on the structure of an individual fragment. These are easily obtainable by substituting into Eq.(5) appropriate elements of matrices $A$ and $C$ (e.g. $A_{ii}=1$ and $C_{ii}=1$ for fragments I and II, respectively) and are collected in Table 2. [Note that $Q^{(1)}=T^{(1)}$ as shown in Eq.(5)].

Table 2
Elements of $\alpha$-dependent matrices $T^{(1)}$, $Q^{(1)}$ and $R^{(1)}$ of Eq.(5) for two-membered conjugated fragments I-IX

| Nr. of fragment | $T_{ii}^{(1)}(Q_{ii}^{(1)})$ | $R_{ii}^{(1)}$ | $T_{ll}^{(1)}(Q_{ll}^{(1)})$ | $R_{ll}^{(1)}$ |
|---|---|---|---|---|
| I | $\alpha/2$ | $\alpha/2$ | 0 | 0 |
| II | $\alpha/2$ | $-\alpha/2$ | 0 | 0 |
| III | $\alpha/2$ | $\alpha/2$ | $\alpha/2$ | $\alpha/2$ |
| IV | $\alpha/2$ | $\alpha/2$ | $\alpha/2$ | $-\alpha/2$ |
| V | $\alpha/2$ | $-\alpha/2$ | $\alpha/2$ | $\alpha/2$ |
| VI | $\alpha$ | 0 | 0 | 0 |
| VII | $\alpha$ | 0 | $\alpha/2$ | $\alpha/2$ |
| VIII | $\alpha$ | 0 | $\alpha/2$ | $-\alpha/2$ |
| IX | $\alpha$ | 0 | $\alpha$ | 0 |

Let us dwell first on fragments I and II containing a single heteroatom (X) in the terminal (1st) position and in the internal (2nd) one, respectively. As is seen from Table 2, the intrabond resonance parameter ($R_{ii}^{(1)}$) between BOs of the Ith double bond ($\phi_{(+)i}$ and $\phi_{(-)i}$) changes its sign when passing from 1- to 2-positioned heteroatom. This seemingly ordinary circumstance, however, yields dramatic consequences upon additional indirect (second order) interactions between BOs.

To discuss this point in a more detail, let us turn to the expression of Eq.(19) for the correction $G_{(2)il}^{(1)}$. Employment of parameters of Rq.(40) along with those of Table 2 shows that the last two contributions of the right-hand side of Eq.(19) vanish for both fragments concerned, whereas the remaining members are of non-zero and uniform absolute values. More importantly, the latter are of the same and of opposite signs for fragments I and II, respectively. In other words, the mediating effects of BOs $\phi_{(+)i}$ and $\phi_{(-)i}$ correspondingly are added together and cancel out one another in the cases I and II. As a result, a non-zero correction $G_{(2)il}^{(1)}$ arises for the first fragment (I) only (Table 1). At the same time, the absolute values of the relevant indirect interactions between BBOs (i.e. of the corrections $D_{(2)il}^{(1+)}$) are uniform for both fragments (Table 1) as turning to Eq.(20) shows [Note that the ABO $\phi_{(-)i}$ is the only mediator in this case]. Besides, an analogous state of things (i.e. addition and cancellation of separate increments to the indirect interaction between a BBO and an ABO along with coinciding interactions between BBOs) has been observed also when comparing linear and cross-conjugated isomers of polyenes [29].

Among further consequences of the distinction between corrections $G_{(2)il}^{(1)}$ (I) and $G_{(2)il}^{(1)}$ (II) there are the relations

$$x_{(+)i,(-)l}^{(3)} = -\frac{\alpha\gamma^2}{8}, \qquad x_{(+)l,(-)i}^{(3)} = \frac{\alpha\gamma^2}{8} \tag{41}$$

and

$$x_{(+)i,(-)l}^{(3)} = x_{(+)l,(-)i}^{(3)} = 0 \tag{42}$$

following for fragments I and II, respectively, after substituting the above-derived elements $G_{(1)il}^{(0)}$ and $G_{(2)il}^{(1)}$ into Eqs.(24) and (25). If we recall Eq.(10) in addition, populations both of the BBO $\phi_{(+)i}$ and of the ABO $\phi_{(-)i}$ are easily seen to grow after introduction of a more electronegative heteroatom (where $\alpha>0$) into the terminal position of butadiene but not into the internal one. In other words, population is expected to be transferred from the Lth (C=C) bond to the Ith (X=C) one in the fragment I. Meanwhile, no charge redistribution among formally-double bonds is predicted to take place in its counterpart II. An analogous qualitative distinction is known between 1- and 2-substituted butadienes in respect of the overall character of redistribution of $\pi$-electrons. In particular, the so-called global and local $\pi$-electron flow, respectively, is ascribed to these isomers [7, 9, 24] on the basis of application of the well-known curly arrow formalism (arrow pushing) [34-36]. Invoking of a certain extended hydrocarbon as a model of the original heteroatom-containing system [7, 9] serves to support this expectation. A more rigorous accounting for the same distinction between substituted isomers follows from graph-theoretical studies of extinction rates of the relevant atom-atom polarizabilities [24].

Let us now return to our fragments I and II to compare the relevant increments to the overall alteration in the CE of a certain compound. The above-established uniform absolute values of corrections $D_{(2)il}^{(1+)}$(I) and $D_{(2)il}^{(1+)}$(II) ensure coinciding negative components $E_{(4-)I-L}^{(2)}$ for fragments I and II. Meanwhile, the relevant positive components $E_{(4+)I-L}^{(2)}$ differ dramatically due to the above-revealed distinction between elements $G_{(2)il}^{(1)}$(I) and $G_{(2)il}^{(1)}$(II). Moreover, the latter (i.e. $E_{(4+)I-L}^{(2)}$) exceeds twice the absolute value of the former (i.e. of $E_{(4-)I-L}^{(2)}$) for the first fragment and, consequently, the total increment $\delta E_{(4)I-L}$(I) takes a positive value. This implies the fragment I to contribute to growth of the CE of the compound concerned. By contrast, the total increment $\delta E_{(4)I-L}$(II) coincides with the negative component $E_{(4-)I-L}^{(2)}$(II). Thus, the fragment II accordingly contributes to reduction of the CE. Again, the above-established positive and negative signs of increments $\delta E_{(4)I-L}$(I) and $\delta E_{(4)I-L}$(II), respectively, indicate a higher overall stability of the 1-substituted butadiene vs. its 2-substituted isomer. [Indeed, both isomers consist of the same formally-double bonds (viz. C=C and C=X) and are consequently represented by uniform local contributions to their total energies ($E^{loc}$ of Eq.(29))]. This outcome, in turn, is in line with the results of the relevant direct HMO calculations [9], as well as with the higher self-polarizability of the terminal carbon atom of butadiene as compared to the internal one (0.63 and 0.40, respectively, in the standard negative units [9, 37]). If we recall finally the decisive role of signs of intrabond resonance parameters ($R_{ii}^{(1)}$) in the formation of both the interbond charge redistribution and relative stabilities of isomers concerned, the above-described distinctions between the latter seem to be of topological origin.

Let us now turn to di-substituted fragments and start with X = C – X = C (III). Since the $2p_z$ AOs of carbon atoms under substitution ($C_i^1$ and $C_l^1$) now belong to the same subset $\{\chi^1\}$, the fragment III is represented by elements $A_{ii}=1$ and $A_{ll}=1$ in the initial Hamiltonian matrix of Eq.(1). The relevant $\alpha$-dependent matrices $T^{(1)}$ ($Q^{(1)}$) and $R^{(1)}$ then contain non-zero elements in both the $i$th and $l$th positions (Table 2). As a result, four non-zero increments of uniform absolute values arise in the expressions of Eq.(19) for $G_{(2)il}^{(1)}$, i.e. four BOs (viz. $\phi_{(+)i}$, $\phi_{(-)i}$, $\phi_{(+)l}$ and $\phi_{(-)l}$) are able to mediate the indirect interaction between BOs $\phi_{(+)i}$ and $\phi_{(-)l}$. Analysis of signs of these increments, however, shows that mediating effects of BOs $\phi_{(+)i}$ and $\phi_{(-)i}$ are added together (as it was the case with the fragment I), whereas the contributions of BOs $\phi_{(+)l}$ and $\phi_{(-)l}$ cancel out one another (in analogy with the fragment II). The ultimate correction $G_{(2)il}^{(1)}$(III) then consequently coincides with that of the fragment I (Table 1). Obviously, the above-discussed additivity of increments of individual mediators to any second order interaction $G_{(2)il}$ plays an important role here. At the same time, the $\alpha$-dependent correction to the indirect interaction between BBOs (i.e. $D_{(2)il}^{(1+)}$ defined by Eq.(20)) vanishes for the fragment III due to cancellation of increments of two mediators ($\phi_{(-)i}$ and $\phi_{(-)l}$).

The above-established coincidence between corrections $G_{(2)il}^{(1)}$(III) and $G_{(2)il}^{(1)}$(I), in turn, implies uniform values both of partial transferred populations $x_{(+)l,(-)i}^{(3)}$ of Eqs.(24) and (25) and of positive contributions to alterations in the CE ($E_{(4+)I-L}^{(2)}$) for fragments III and I (Table 1). More surprisingly, the relevant negative increments $E_{(4-)I-L}^{(2)}$(III) and $E_{(4-)I-L}^{(2)}$(I) also take the same values in spite of entirely different corrections $D_{(2)il}^{(1+)}$ (III) and $D_{(2)il}^{(1+)}$ (I). The reason lies in growth of the second term of Eq.(38) due to the increasing number of C=X bonds ($n_{C=X}$) from 1 to 2 when passing from I to III. Consequently, the relevant total contributions also are uniform, i.e. $\delta E_{(4)I-L}$(III) coincides with $\delta E_{(4)I-L}$(I). This outcome, in turn, indicates the overall effect of two heteroatoms to be of a non-additive nature inside the di-substituted fragment III. As opposed to the above-compared fragments I and II, however, the relevant local contributions of Eq.(29) (i.e $E^{loc}$(I) and $E^{loc}$(III)) differ one from another. Thus, the above-established uniform values of increments $\delta E_{(4)I-L}$(III) and $\delta E_{(4)I-L}$(I) do not ensure coinciding overall stabilities of respective mono- and di-substituted derivatives of butadiene.

Let us now dwell on the subsequent di-substituted fragments IV and V, where heteroatoms stand for carbon atoms, the $2p_z$ AOs of which belong to different subsets. As a result, elements $A_{ii}=C_{ll}=1$ and $C_{ii}=A_{ll}=1$ represent fragments IV and V, respectively, in the initial Hamiltonian matrix of Eq.(1). As with the former fragment III, all the four BOs are then among potential mediators of indirect interactions $G_{(2)il}^{(1)}$(IV) and $G_{(2)il}^{(1)}$(V). Actually, however, there are two positive and two negative contributions in the relevant expressions of Eq.(19), e.g. positive increments of BOs $\phi_{(+)i}$ and $\phi_{(-)i}$ to $G_{(2)il}^{(1)}$(IV) go together with negative ones of BOs $\phi_{(+)l}$ and $\phi_{(-)l}$. Consequently, the corrections $G_{(2)il}^{(1)}$ vanish for both fragments concerned along with related characteristics, viz. partial transferred populations ($x_{(+)l,(-)i}^{(3)}$) and positive components of energetic increments ($E_{(4+)I-L}^{(2)}$). By contrast, contributions of both mediators (i.e. of ABOs $\phi_{(-)i}$ and $\phi_{(-)l}$) are added together in the expressions of Eq.(20) for corrections $D_{(2)il}^{(1+)}$ (IV) and $D_{(2)il}^{(1+)}$ (V) so that these consequently take double higher absolute values as compared to $D_{(2)il}^{(1+)}$ (I) and $D_{(2)il}^{(1+)}$ (II). Moreover, absolute values of related energetic increments $E_{(4-)I-L}^{(2)}$ (IV) and $E_{(4-)I-L}^{(2)}$ (V) are even three times higher vs. those of $E_{(4-)I-L}^{(2)}$ (I) and $E_{(4-)I-L}^{(2)}$ (II) due to passing from $n_{C=X}=1$ to $n_{C=X}=2$. As a result, fragments IV and V are characterized by negative total contributions to the overall alteration in the CE of relatively high absolute value. Non-additivity of the total effect of two heteroatoms also deserves mention here.

Let us now pay some attention to comparison of fragments III, IV and V consisting of two C=X bonds and consequently represented by uniform local increments to their total energies ($E^{loc}$ of Eq.(29)). The above-established positive sign of the increment $\delta E_{(4)I-L}$(III) along with negative ones of the remaining contributions ($\delta E_{(4)I-L}$(IV) and $\delta E_{(4)I-L}$(V)) then indicate a higher overall stability of the 1,3-di-substituted butadiene as compared to its 1,4- and 2,3-di-substituted isomers. Analogous predictions follow also from graph-theoretical studies of alternant systems with two heteroatoms [21], as well as from application of the CLHPT. In the latter case, the total alteration in the $\pi$-electron energy of an $r,s$-di-substituted compound depends on combination of self- and mutual polarizabilities of the relevant carbon atoms, viz. $\pi_{rr}+\pi_{ss}+2\pi_{rs}$. After substituting the respective numerical values in the standard negative units [37], the above-specified decisive combinations coincide with 1.118, 0.716 and 0.714 for 1,3-, 1,4- and 2,3-disubstituted butadienes, respectively, and point to destabilization of the last two isomers vs. the first one.

As opposed to the above-considered di-substituted fragments III-V, the last one (VI) contains an X=X bond, the latter being represented by a zero intrabond resonance parameter $R_{ii}^{(1)}$ (Table 2) due to its homopolar nature. Meanwhile, elements of the remaining $\alpha$-dependent matrices of Eq.(5) (i.e. of $\boldsymbol{T}^{(1)}$ and $\boldsymbol{Q}^{(1)}$) take double values ($\alpha$) vs. those associated with C=X bonds. Turning to Eq.(19) then shows that the BBO $\phi_{(+)i}$ actually is the only mediator of the indirect interaction $G_{(2)il}^{(1)}$(VI) and the latter ultimately coincides with the relevant corrections for fragments I and III. An analogous coincidence consequently refers also to partial transferred populations and to positive components of the relevant energetic increments. Again, employment of the equality $R_{ii}^{(1)}=0$ yields a vanishing correction $D_{(2)il}^{(1+)}$(VI) and, consequently, a zero component $E_{(4-)I-L}^{(2)}$(VI) [Note that $n_{C=X}=0$ for the fragment concerned]. The total contribution of our fragment to the overall alteration in the CE ($\delta E_{(4)I-L}$(VI)) then coincides with its positive component $E_{(4+)I-L}^{(2)}$ (VI) and exceeds both $\delta E_{(4)I-L}$ (I) and $\delta E_{(4)I-L}$ (III). Thus, the fragment VI contributes significantly to an increase in the CE of the relevant compound.

Fragments of Table 1 with three heteroatoms (VII and VIII) offer no new aspects to the present discussion due to their similarity to mono-substituted ones (i.e. to I and II, respectively). This result causes little surprise if we observe that the only carbon atom of fragments VII and VIII plays the role of the heteroatom here. Accordingly, the last fragment (IX) is an analogue of the parent butadiene.

It is seen, therefore, that the decisive indirect interaction between the BBO $\phi_{(+)i}$ and the ABO $\phi_{(-)l}$ ($G_{(2)il}^{(1)}$) meets a simple selection rule for above-considered fragments I-IX, viz. it coincides with a constant ($\alpha\gamma/8$) for fragments I, III, VI and VII and takes a zero value for the remaining ones (II, IV, V, VIII and IX). An analogous rule then consequently refers both to partial populations transferred inside fragments ($x_{(+)l,(-)i}^{(3)}$) and to positive components of energetic increments ($E_{(4+)I-L}^{(2)}$). Due to predominance of the latter over their negative counterparts, fragments of the first group are ultimately characterized by positive total contributions to the overall alteration in the CE and denoted by A in the last column of Table 1. Meanwhile, fragments of the second group (B) offer negative (or zero) energetic increments $\delta E_{(4)I-L}$.

## Conclusion

1. Application of the perturbational model to acyclic polyenes with heteroatom(s) yields expressions of a high degree of additivity for both the relevant total $\pi$-electron energy and populations of basis orbitals. First of all, local and transferable increments of individual formally-double bonds are noteworthy that support the common image of compounds concerned as consisting of weakly-interacting C=C, C=X and X=X bonds. Further, additive and transferable contributions of less local nature deserve attention that originate from separate conjugated substructures (I–L) consisting of two connected formally-double bonds (I and L) and containing at least a single heteroatom. At the same time, these contributions are responsible for most parts of both the overall alteration in the CE of the given compound due to introduction of heteroatom(s) and the relevant charge redistribution among formally-double bonds. Hence, substructures

I–L actually play the role of elementary conjugated fragments of polyenes with heteroatom(s).
2. Nine potential elementary fragments (I-IX) reveal themselves in the systems concerned that contain up to four heteroatoms in different positions and generally make distinct contributions to the above-specified characteristics of the whole compound. So far as increments of individual heteroatoms inside the same fragment are concerned, these are neither additive nor transferable. Thus, the above-concluded global additivity of heteroatom influence goes together with its local non-additivity.
3. The nine potential elementary fragments (I-IX) may be further classified into two groups. The first one embraces fragments I, III, VI and VII characterized by positive contributions to the overall alteration in the CE due to introduction of heteroatom(s) and by non-zero increments to the relevant charge redistribution among formally-double bonds. Meanwhile, negative and zero contributions, respectively, are peculiar to fragments of the second group (viz. II, IV, V, VIII and IX).
4. The perturbative approach applied enables us also to represent heteroatom influence in terms of additional interactions between BOs of C=C bonds of the parent polyene that undergo a certain gradual expansion over the carbon backbone: At the early stage, introduction of one or two heteroatoms into the Ith C=C bond of the parent polyene is accompanied by emergence of direct intrabond self- and mutual interactions of BOs of only this bond. Thereupon, these newly-formed local interactions potentially give birth to long-range interactions of various types. As the most important examples of the latter, indirect interactions of BOs of two principal types deserve mention that embrace the nearest neighbourhood of the Ith double bond (i.e. all elementary conjugated fragments I–L, the Ith bond participates in), viz. (i) stabilizing indirect interactions between the BBO of the Ith bond and the ABO of the Lth one (or vice versa) that determine the relevant interbond charge redistribution in addition and (ii) destabilizing indirect interactions between BBOs of the same double bonds.
5. Analysis of expressions for the above-specified indirect interorbital interactions provides us with a rationale for existance of two discrete groups of elementary conjugated fragments. The point is that the decisive stabilizing interactions actually take non-zero and zero values for fragments of the first group and those of the second one, respectively, cancellation of increments of individual mediators of this interaction being the reason in the latter case. Just this fact forms the basis for a clear distinction beween characteristics of fragments belonging to different groups.

## Methods

*The essentials of the PNCMO theory*
Let us start with the well-known PMO theory [38] that is commonly understood as a combination of the Hückel model [31, 32] and of the standard Rayleigh-Schrödinger PT (RSPT) (see e.g. [39]). The eigenvalue equation for the Hamiltonian matrix of a certain molecule is solvable by means of this theory under condition of significant energy gap(s) vs. the interorbital interaction(s) (resonance parameter(s)). Eigenvectors and eigenvalues of this matrix are then obtainable that coincide with respective canonical MOs (CMOs) and their energies. The latter, in turn, form the basis for a subsequent derivation of related characteristics including the charge-bond order (CBO) matrix and total energy.

Let us now address a simple two-level two-electron system under the above-mentioned condition. The only interorbital interaction (resonance parameter) is assumed to vanish in the relevant initial (zero order) approximation. The orbitals will be denoted by $\phi_{(+)}$ and $\phi_{(-)}$, where the subscripts (+) and (–) indicate the orbital concerned to be initially double-occupied and vacant, respectively. The energy reference point will be chosen in the middle of the interorbital energy gap and a negative energy unit will be used in addition. The initial (zero order) energies of our orbitals will be correspondingly designated by $\varepsilon_{(+)}$ and $-\varepsilon_{(-)}$, where $\varepsilon_{(+)} > 0$ and $\varepsilon_{(-)} > 0$.

Let the overall perturbation of our system embrace both the only resonance parameter (denoted by $\rho$) and certain alterations in one-electron energies of orbitals $\phi_{(+)}$ and $\phi_{(-)}$ themselves (correspondingly designated by $\tau$ and $\kappa$). Moreover, parameters $\rho$, $\tau$ and $\kappa$ are supposed to be first order terms vs. the energy gap $\varepsilon_{(+)} + \varepsilon_{(-)}$. The total Hamiltonian matrix of our system ($h$) is then representable as a sum of the zero order member ($h_{(0)}$) and of the first order one ($h_{(1)}$) as follows

$$h = h_{(0)} + h_{(1)} = \begin{vmatrix} \varepsilon_{(+)} & 0 \\ 0 & -\varepsilon_{(-)} \end{vmatrix} + \begin{vmatrix} \tau & \rho \\ \rho & \kappa \end{vmatrix}. \tag{43}$$

Obviously, the eigenvalue equation for the matrix $h$ is easily solvable by means of the RSPT and two MOs ultimately result along with their actual energies $\varepsilon_1$ and $\varepsilon_2$.

Let us now introduce an important generalization of the above-described simple system, where certain sets of basis functions $\{\phi_{(+)}\}$ and $\{\phi_{(-)}\}$ play the role of the former orbitals $\phi_{(+)}$ and $\phi_{(-)}$, respectively. As a result, we turn to molecules and/or molecular systems representable by two weakly-interacting well-separated subsets of basis functions. Further, the subsets $\{\phi_{(+)}\}$ and $\{\phi_{(-)}\}$ are assumed to contain $n$ and $m$ orbitals, respectively, that are correspondingly supposed to be double-occupied and vacant before perturbation. [Note that $n$ does not necessarily coincide with $m$]. The most general form of the relevant Hamiltonian matrix then resembles that of Eq.(43) except for submatrices (blocks) playing the role of the former parameters (matrix elements), viz.

$$\bar{H} = \bar{H}_{(0)} + \bar{H}_{(1)} = \begin{vmatrix} E_{(+)} & 0 \\ 0 & -E_{(-)} \end{vmatrix} + \begin{vmatrix} T & R \\ R^* & Q \end{vmatrix}. \tag{44}$$

Submatrices $E_{(+)} + T$ and $-E_{(-)} + Q$ now contain the Hamiltonian matrix elements between orbitals inside subsets $\{\phi_{(+)}\}$ and $\{\phi_{(-)}\}$, respectively, whereas $R$ embraces the intersubset interactions. Besides, intrasubset interactions of the zero order magnitude are allowed in the matrix $\bar{H}$ and these coincide with off-diagonal elements of submatrices $E_{(+)}$ and $E_{(-)}$. It also deserves emphasizing that neither the internal constitutions of submatrices nor their dimensions ($n$ and $m$) are specified in the matrix $\bar{H}$. Finally, comparison of the matrix $\bar{H}$ to $H'$ of Eq.(3) shows the latter to coincide with a particular case of the former, where $E_{(+)}=E_{(-)}=I$ and $n=m$. Thus, polyenes with heteroatom(s) offer us an example of systems representable by the matrix $\bar{H}$.

The non-trivial nature of the above-introduced generalization (i.e. passing from the matrix $h$ of Eq.(43) to $\bar{H}$ of Eq.(44)) is evident. First, the latter represents a wide class of molecules and molecular systems in contrast to usual Hamiltonian matrices of individual molecules. Second, solution of the standard (canonical) one-electron problem (i.e. of the eigenvalue equation) is no longer possible for the matrix $\bar{H}$ [40]. Thus, we necessarily have to turn to alternative (non-canonical) problems.

As is known, the eigenvalue equation for any matrix is equivalent to the respective diagonalization problem. In particular, a diagonal form embracing two eigenvalues $\varepsilon_1$ and $\varepsilon_2$ is sought in the case of the matrix $h$ of Eq.(43). A direct generalization of this procedure to the matrix $\bar{H}$ of Eq.(44) accordingly consists in search for a block-diagonal form of the latter coinciding with a direct sum of two submatrices (blocks) $E_1$ and $E_2$ associated with subspaces of double-occupied and vacant orbitals, respectively, and called the eigenblocks (see e.g. [30, 41] and the references cited therein). At the same time, the block-diagonalization problem is among particular forms of the Brillouin theorem [42, 43] determining the relevant non-canonical (localized) MOs (NCMOs). [The latter are contained within the respective transformation matrix]. It is then no surprise that just the above-described option deserves our attention in the first place.

Solution of the block-diagonalization problem for the matrix $\bar{H}$ [41, 44] of Eq.(44) was based on (i) search for both the eigenblocks ($E_1$ and $E_2$) and the transformation matrix ($C$) in the form of power series with respect to submatrices $T$, $Q$ and $R$ and (ii) on an initial representation of the matrix $C$ in terms of four submatrices (blocks) of appropriate dimensions (coinciding with those of blocks of the matrix $\bar{H}$). Meanwhile, submatrices themselves (including those of Eq.(44) and the blocks being sought) were treated as "indivisible" quantities. As a result, the solution concerned has been expressed in terms of entire submatrices of the matrix $\bar{H}$ ($E_{(+)}$, $E_{(-)}$, $T$, $Q$ and $R$) without specifying either constitutions or dimensions of the latter. Since submatrices generally are non-commutative quantities (in contrast to usual matrix elements), the newly-elaborated perturbative approach has been accordingly called the non-commutative Rayleigh-Schrödinger PT (NCRSPT) [45, 46]. Derivation of the relevant CBO matrix ($P$) and of the total energy ($E$) was the second step of the overall procedure as usual. In particular, the matrix $P$ proved to be representable via entire submatrices of the transformation matrix ($C$) [41, 46] in analogy with the CBO matrix of our two-orbital system ($p$) following from elements of the 2x2-dimensional matrix of eigenvectors of the matrix $h$. As a result, the matrix $P$ also has been ultimately expressed via entire blocks of the matrix $\bar{H}$. Meanwhile, the relevant total energy ($E$) followed from the two-fold trace of the occupied eigenblock $E_1$ [47]. Just this scheme has been originally called the perturbational NCMO (PNCMO) theory [41, 46].

The above-outlined procedure, however, proved to be not the optimum one in practice, especially if NCMOs themselves are not among characteristics under our interest. Indeed, the overall derivation of the CBO matrix $P$ and/or of the total energy $E$ is rather cumbersome in this case [47], in particular because of ambiguity of the transformation matrix (NCMOs) [44]. That is why the less-known direct way of derivation of any CBO matrix [48] also has been invoked when constructing the PNCMO theory. The underlying non-canonical problem then embraces three matrix equations exhibited below and determining the matrix $P$ without any reference to either CMOs or NCMOs. In contrast to the above-described passing from the eigenvalue equation to the eigenblock problem accompanied by an increasing extent of ambiguity of solutions, the direct way of derivation of the CBO matrix is straightforwardly extendable to systems representable by the matrix $\bar{H}$ of Eq.(44). The relevant system of equations to be solved is then as follows

$$[\bar{H}, P]_- = 0, \quad P^2 = 2P, \quad \mathrm{Tr}\, P = 2n, \qquad (45)$$

where the notation $[\ldots,\ldots]_-$ stands for a commutator of matrices. The first relation of Eq.(45) (the commutation condition) is the main physical requirement determining the matrix $P$ and resulting from the Dirac equation for the time-independent Hamiltonian [48]. Meanwhile, the remaining relations of Eq.(45) are additional system-structure-independent restrictions following from the idempotence requirement ($\Pi^2 = \Pi$) for the projector $\Pi = P/2$ and from the charge conservation condition, respectively.

The principal points underlying the solution of equations of Eq.(45) [44] resemble those of the block-diagonalization problem: First, the matrix $P$ has been sought as a sum of increments $P_{(k)}$ of increasing orders ($k$) with respect to blocks of the first order Hamiltonian matrix ($T$, $Q$ and $R$), viz.

$$P = \sum_{k=0}^{\infty} P_{(k)} = P_{(0)} + P_{(1)} + P_{(2)} + ... \qquad (46)$$

Second, both the total matrix $P$ and its individual increments $P_{(k)}$, $k$=0, 1, 2… have been initially represented via four submatrices (blocks). Consequently, separate blocks of the matrix $P$ also take the form of power series like that of Eq.(46). Finally, the above-specified overall form of the matrix $P$ has been substituted into relations of Eq.(45) and terms of each order ($k$) separately have been collected (as is usual in perturbative approaches). Let us now turn to the relevant results.

The zero order member of the expansion of Eq.(46) is as follows

$$P_{(0)} = 2 \begin{vmatrix} I & 0 \\ 0 & 0 \end{vmatrix} \qquad (47)$$

and represents the initial occupation numbers of basis orbitals coinding with 2 and 0 for those of subsets $\{\phi_{(+)}\}$ and $\{\phi_{(-)}\}$, respectively. Meanwhile, the remaining terms of the same series ($P_{(k)}$, $k$=1, 2…) take the following form

$$P_{(k)} = -2 \begin{vmatrix} D_{(k)}^{(+)} & G_{(k)} \\ G_{(k)}^* & -D_{(k)}^{(-)} \end{vmatrix}, \qquad (48)$$

where the front factor (–2) is introduced for convenience. Matrices $D_{(k)}^{(+)}$ and $D_{(k)}^{(-)}$ are expressible via matrices $G_{(k)}$, $k$=1,2… as shown in Eq.(6) and are correspondingly associated with subsets of basis orbitals $\{\phi_{(+)}\}$ and $\{\phi_{(-)}\}$ as their positions in the matrices $P_{(k)}$ indicate. The principal matrices $G_{(k)}$, $k$=1,2…, in turn, meet the following matrix equations

$$E_{(+)} G_{(k)} + G_{(k)} E_{(-)} + W_{(k)} = 0 , \qquad (49)$$

where

$$W_{(1)} = R, \qquad W_{(2)} = T G_{(1)} - G_{(1)} Q,$$
$$W_{(3)} = T G_{(2)} - G_{(2)} Q - (R G_{(1)}^* G_{(1)} + G_{(1)} G_{(1)}^* R), etc. \qquad (50)$$

Consequently, the actual occupation numbers (populations) of individual basis orbitals follow from respective diagonal elements of submatrices taking diagonal positions within contributions $P_{(0)}$ and $P_{(k)}$ of Eqs.(47) and (48) after summing them over the order parameter $k$ in accordance with Eq.(46). For orbitals $\phi_{(+)i}$ and $\phi_{(-)l}$ of subsets $\{\phi_{(+)}\}$ and $\{\phi_{(-)}\}$, respectively, expressions of Eqs.(8)-(11) are then easily obtainable.

Furthermore, the relevant total energy ($E$) also has been alternatively derived directly from the matrix $P$ using the following general relation [48]

$$E = Tr(P \bar{H}). \qquad (51)$$

Substituting Eqs.(44) and (46)-(48) into Eq.(51) shows that the total energy $E$ also takes the form of an analogous power series. An additional employment of Eqs.(49) and (50) then yields the following expressions for starting members of this expansion [49]

$$E_{(0)} = 2 Tr E_{(+)}, \qquad E_{(1)} = 2 Tr T, \qquad E_{(2)} = -2 Tr(G_{(1)} R^*), \qquad E_{(3)} = -2 Tr(G_{(2)} R^*),$$
$$E_{(4)} = -2 Tr[(G_{(3)} + G_{(1)} G_{(1)}^* G_{(1)}) R^*]. \qquad (52)$$

It is seen, therefore, that two ways of derivation of many-electron characteristics ($P$ and $E$) exist for systems representable by the Hamiltonian matrix of Eq.(44) that evidently yield coinciding results. Moreover, the underlying non-canonical problems themselves (i.e. the block-diagonalization problem and matrix equations of Eq.(45)) are deeply interelated in this case as the relevant analysis shows [44]. Thus, it is quite natural to consider these alternative schemes as parts of the same PNCMO theory.

As already mentioned, the particular case $\boldsymbol{E}_{(+)}=\boldsymbol{E}_{(-)}=\boldsymbol{I}$ inside Eq.(44) corresponds to our polyenes with heteroatom(s). Matrix equations of Eq.(49) are then solvable algebraically and yield expressions for matrices $\boldsymbol{G}_{(k)}$, $k=1,2\ldots$ shown in Eq.(7). Further, replacement of the matrix $\boldsymbol{R}^*$ of Eq.(52) by $-2\boldsymbol{G}^*_{(k)}$ in accordance with the first relation of Eq.(7) results into formulae for corrections $E_{(2)}$ and $E_{(3)}$ in terms of matrices $\boldsymbol{G}_{(1)}$ and $\boldsymbol{G}_{(2)}$ exhibited in Eq.(12). The respective fourth order term ($E_{(4)}$) is then accordingly expressible via the matrix $\boldsymbol{G}_{(3)}$. Passing to expressions of Eqs.(13) and (14) [33] was based on employment of the last relation of Eq.(50).

**References**
1. Trinajstić N (2018) Chemical graph theory. CRC Press, Boca Raton, FL
2. Bonchev D, Rouvray DH (1991) Chemical graph theory: Introduction and fundamentals. CRC Press, Boca Raton, FL
3. Trinajstić N (1977) In: Segal GA (ed) Semiempirical methods of electronic structure calculations, Part A, Techniques. Plenum Press, New York and London
4. Gutman I (2005) J Serb Chem Soc 70:441
5. Gutman I, Furtula B (2017) Croat Chem Acta 90:359
6. Debska B, Guzowska-Swider B (2000) J Chem Comput Sci 40:325
7. Hosoya H (2015) Curr Org Chem 19:293
8. Hosoya H (2017) Molecules 22:896
9. Hosoya H (2003) Bull Chem Soc Jpn 76:2233
10. Dias JR (2014) J Phys Chem A118:10822
11. Dias JR (2016) Polycyclic Arom Comp 36:544
12. Randić M (2003) Chem Rev 103:3449
13. Gutman I (2005) Monatsh Chem 136:1055
14. Gutman I, Ruščić B, Trinajstić N, Wilcox CF Jr (1975) J Chem Phys 62:3399
15. Dias JR (1993) Molecular orbital calculations using chemical graph theory. Springer-Verlag, Berlin
16. Dias JR (1987) MATCH Commun Math Comput Chem 22:257
17. Dias JR (1997) J Phys Chem A 101:7167
18. Gutman I, Bosanac S (1976) Chem Phys Lett 43:371
19. Gutman I (1979) Theor Chim Acta 50:287
20. Gutman I (1981) Z Naturforsch A 36:1112
21. Gutman I (1990) Z Naturforsch A 45:1085
22. Coulson CA, Longuet-Higgins HC (1947) Proc Roy Soc (London) A 191:39
23. Coulson CA, Longuet-Higgins HC (1947) Proc Roy Soc (London) A 192:16
24. Hosoya H (1999) J Mol Struct (Theochem) 461-462:473
25. Vilkov LV, Mastryukov VS, Sadova NI (1983) Determination of the geometrical structure of free molecules, Mir Publishers, Moscow
26. Milián-Medina B, Gierschner J (2012) WIREs Comput Mol Sci 2:513
27. Craig NC, Groner P, McKean DC (2006) J Phys Chem A110:7461
28. Schmalz TG, Griffin LL (2009) J Chem Phys 131:224301
29. Gineityte V (2016) Monatsh Chem 147:1303
30. Gineityte V (2020) Monatsh Chem 151:899
31. Coulson CA, O'Leary B, Mallion RB (1978) Hűckel theory for organic chemists, Academic Press, London
32. Yates K (1978) Hűckel molecular orbital theory. Acad Press, New York
33. Gineityte V (2013) Int J Chem Model 5:99

34. O'Hagan D, Lloyd D (2010) Chem World 7:54
35. Levy DE (2008) Arrow-pushing in organic chemistry: An easy approach to understanding reaction mechanisms, Wiley, New York
36. Andres J, Berski S, Silvi B (2016) Chem Commun 52:8183
37. Basilevskii MV (1969) Metod molekuliarnych orbitalei i reaktsionnaya sposobnost organitcheskich molecul, Khimia, Moscow
38. Dewar MJS, Dougherty RC (1975) The PMO theory in organic chemistry. Plenum Press, New York
39. Zülicke L (1973) Quantenchemie, bd 1, Grundlagen and algemeine methoden, VEB Deutcher Verlag der Wissenschaften, Berlin
40. Gineityte V (1995) J Mol Struct (Theochem) 333:297
41. Gineityte V (2018) ArXiv:1803.03156. http://arxiv.org/abs/1803.03156
42. S. Huzinaga (1992) Molecular orbital method, Cambridge Univ Press, Cambridge
43. Chalvet O, Daudel R, Diner S, Malrieu JP (eds) (1975) Localization and delocalization in quantum chemistry. v 1, Atoms and molecules in the ground state. Reidel, Dordrecht
44. Gineityte V (1995) J Mol Struct (Theochem) 343:183
45. Gineityte V (1998) Int J Quant Chem 68:119
46. Gineityte V (2004) Lith J Phys 44:219
47. Gineityte V (2011) Lith J Phys 51:107
48. Mestetchkin (1977) The density matrix method in quantum chemistry, Naukova Dumka, Kiev
49. Gineityte V (2002) J Mol Struct (Theochem) 585:15